\documentclass[useAMS,usenatbib]{mn2e}
\usepackage[]{natbib}
\usepackage{amsmath} 
\usepackage{amssymb} 
\usepackage{graphicx} 
\usepackage{color} 
\newcommand{\HI}{\mbox{${\rm H\,I}$}}

\newcommand{\MgII}{{\mbox{Mg\,{\scriptsize II}\ }}}
\newcommand{\CIV}{{\mbox{C\,{\scriptsize IV}\ }}}

\newcommand{\OII}{{\mbox{[O\,{\scriptsize II}]}}}
\newcommand{\NII}{{\mbox{[N\,{\scriptsize II}]}}}
\newcommand{\OIII}{{\mbox{[O\,{\scriptsize III}]}}}
\newcommand{\Halpha}{{\mbox{H$\alpha$}}}
\newcommand{\Hdelta}{{\mbox{H$\delta$}}}

\newcommand{\apg}{\gtrsim}
\newcommand{\apl}{\lesssim}
\newcommand{\etal}{{\mbox{et~al.}}}

\newcommand{\ewr}{\mbox{$W_r(2796)$}}
\newcommand{\kms}{\,${\rm{km\,s}^{-1}}$}

\newcommand{\mnras}{MNRAS} \newcommand{\aj}{AJ}
\newcommand{\apj}{ApJ} \newcommand{\apjl}{ApJL}
\newcommand{\apjs}{ApJS} \newcommand{\aap}{A\&A}
\newcommand{\araa}{ARA\&A} \newcommand{\jcap}{JCAP}
\newcommand{\nat}{Nature}

\title[Chemically-enriched CGM around 38000 LRGs]{Characterizing the Chemically-Enriched Circumgalactic Medium of $\sim$ 38000 Luminous Red Galaxies in SDSS DR12}

\author[Huang \etal]{Yun-Hsin Huang$^{1}$\thanks{E-mail: yunhsin@uchicago.edu}, Hsiao-Wen Chen$^{1}$\thanks{E-mail: 
hchen@oddjob.uchicago.edu}, Sean D.\ Johnson$^{1}$ and Benjamin J.\ Weiner$^{2}$\\
$^{1}$Department of Astronomy \& Astrophysics, and Kavli Institute for 
Cosmological Physics, The University of Chicago, Chicago, IL 60637, USA\\
$^{2}$Steward Observatory, University of Arizona, Tucson, AZ 85721, USA\\ }
\begin{document}

\date{\today}

\pagerange{\pageref{firstpage}--\pageref{lastpage}} \pubyear{2015}

\hsize=6truein
\maketitle

\label{firstpage}

\begin{abstract}

  We report a definitive detection of chemically-enriched cool gas
  around massive, quiescent galaxies at $z\approx 0.4-0.7$.  The
  result is based on a survey of 37621 luminous red galaxy (LRG)-QSO
  pairs in SDSS DR12 with projected distance $d<500$ kpc.
%Of these LRGs, 13330 have sensitive background QSO spectra
%  available for constraining the presence/absence of \MgII absorbers
%  of rest-frame absorption equivalent width $\ewr\ge 0.3$ \AA.
  The LRGs are characterized by a predominantly old (${\rm
    age}\,\apg\,1$ Gyr) stellar population with 13\% displaying
  \OII\ emission features and LINER-like spectra.  Both passive and
  \OII-emitting LRGs share the same stellar mass distribution with a
  mean of $\langle\,\log\,(M_*/M_\odot)\,\rangle \approx 11.4$ and a
  dispersion of 0.2 dex.  Both LRG populations exhibit associated
  strong \MgII absorbers out to $d<500$ kpc.  The mean gas covering
  fraction at $d\apl 120$ kpc is $\langle\kappa\rangle_\MgII> 15$\%
  and declines quickly to $\langle\kappa\rangle_\MgII\approx 5$\% at
  $d\apl 500$ kpc.
%While the mean
%  absorber strength of detections remains flat out to the virial
%  radius, $\langle\kappa\rangle_\MgII$ decreases rapidly with
%  increasing distance.  Specifically, we find
%  $\langle\kappa\rangle_\MgII> 15$\% at $d\apl 120$ kpc and
%  $\langle\kappa\rangle_\MgII\approx 11-14$\% at $d< 200$ kpc.
%  At
%  $d<80$ kpc, \OII-emitting LRGs display higher
%  $\langle\kappa\rangle_\MgII$ than passive LRGs by a factor of
%  $\approx 2$.  
No clear dependence on stellar mass is detected for
  the observed \MgII absorption properties.  The observed velocity
  dispersion of Mg\,II absorbing gas relative to either passive or
  \OII-emitting LRGs is merely 60\% of what is expected from virial
  motion in these massive halos.  While no apparent azimuthal
  dependence is seen for $\langle\kappa\rangle_\MgII$ around passive
  LRGs at all radii, a modest enhancement in
  $\langle\kappa\rangle_\MgII$ is detected along the major axis of
  \OII-emitting LRGs at $d<50$ kpc.  The suppressed velocity
  dispersion of \MgII absorbing gas around both passive and
  \OII-emitting LRGs, together with an elevated
  $\langle\kappa\rangle_\MgII$ along the major axis of \OII-emitting
  LRGs at $d<50$ kpc, provides important insights into the origin of
  the observed chemically-enriched cool gas in LRG halos.  We consider
  different scenarios and conclude that the observed \MgII absorbers
  around LRGs are best-explained by a combination of cool clouds
  formed in thermally unstable LRG halos and satellite accretion
  through filaments.

\end{abstract}


\section{Introduction}
\label{section:introduction}

Luminous red galaxies (LRGs) uncovered in the Sloan Digital Sky Survey
(SDSS; York \etal\ 2000; Eisenstein \etal\ 2011) have luminosities of
$\approx 5\,L_*$ \citep[e.g. ][]{Tojeiro:2011} and reside in halos of
$M_{\rm halo} \apg 10^{13} M_\odot$
\citep[e.g.,][]{Blake:2008,Zhu:2014}.  These galaxies exhibit optical
colors that resemble nearby elliptical galaxies with little on-going
star formation \citep{Eisenstein:2001}.  The LRGs constitute a
homogeneous sample of massive galaxies characterized by old stellar
populations and provide an ideal laboratory for studying galaxy
formation and evolution at the high-mass end.

\MgII $\lambda\lambda$ 2793,2803 absorption features are commonly seen
in the spectra of distant QSOs \citep[e.g.,][]{Charlton:2003}.  These
absorbers originate in cool gas of temperature $T\sim10^4\,$K
\citep[e.g.,][]{Bergeron:1986} and neutral hydrogen column density
ranging from $N{(\rm H\,I)}\apl 10^{18}\, {\rm cm}^{-2}$ to $N({\rm
  H\,I)}\approx 10^{22}\, {\rm cm}^{-2}$ \cite[e.g.,][]{Rao:2006}, and
provide a sensitive probe of diffuse interstellar medium (ISM) and
circumgalactic medium (CGM) along individual QSO sightlines.

It has been well-established over the past two decades that typical
$L_*$ and sub-$L_*$ galaxies are surrounded by extended \MgII
absorbing gas out to projected distance of $d=50-100$ kpc with a mean
gas covering fraction of $\langle\kappa\rangle_\MgII\apg 70$\% (e.g.,
Bowen \etal\ 1995; Chen \& Tinker 2008; Chen \etal\ 2010a)\footnote{We
  note a critical distinction between \MgII-selected galaxy surveys
  and galaxy-centric absorber searches.  The former, \MgII-selected
  galaxy surveys, address questions regarding the origin of detected
  \MgII absorbers (e.g., Steidel \etal\ 1994, 1997), while the latter,
  galaxy-centric absorber searches, address the incidence and covering
  fraction of \MgII absorbing gas around known galaxies.  Including
  \MgII-absorbing galaxies uncovered from absorber-selected studies
  would naturally introduce significant scatter and bias the observed
  gas covering fraction to higher values.}.  The presence of
chemically-enriched cool gas at $\sim 100$ kpc from star-forming
regions can be explained by two competing models: infall (e.g., Mo \&
Miralda-Escud\'e 1996; Maller \& Bullock 2004) and outflows (e.g.,
Weiner \etal\ 2009; Murray \etal\ 2011).  Depending on the relative
locations of the absorber and its associated galaxy, infalling gas and
outflowing material can show similar line-of-sight velocity offset.
In addition, depending on how efficiently chemical species are mixed
in with surrounding medium, outflowing materials can appear as
low-metallicity gas.  Consequently, distinguishing between the two
scenarios based on available empirical data remains challenging.

CGM studies focusing on halos around quiescent galaxies offer a
promising avenue for resolving the ambiguous contributions between
infall and outflows to the observed absorber statistics in the halos.
The predominantly old stellar populations of quiescent galaxies,
together with little/no on-going star formation, indicate a
diminishing influence of young starburst driven winds on their halo
gas.  \MgII absorbers have been found to cluster strongly with LRGs,
which suggests a non-negligible presence of these \MgII absorbers in
these massive, quiescent halos
\citep[e.g.,][]{Gauthier:2009,Lundgren:2009,Zhu:2014}.  Spectroscopic
follow-up of close LRG-QSO pairs by \cite{Gauthier:2010} and
\cite{Gauthier:2011} further confirmed that indeed roughly $(14\pm
6)$\% of LRGs have associated \MgII absorbers at $d<500$ kpc.
%, but with 43\%
%uncertainties were large $\approx 43$\%.

The non-negligible presence of chemically-enriched cool gas near LRGs
extends the findings of \HI\ gas in nearby elliptical galaxies
\citep[e.g.,][]{Oosterloo:2010} to higher redshifts.  These massive,
quiescent galaxies provide an ideal laboratory for testing possible
physical mechanisms to widely distribute heavy elements away from
stars in the absence of starburst outflows.  Likely scenarios include
cold flows or filaments from the intergalactic medium
\citep[e.g.,][]{Keres:2009, Faucher:2011, Nelson:2013},
pressure-supported cool clouds in a hot halo
\citep[e.g.,][]{Mo:1996,Maller:2004}, and stripped materials from
satellite galaxies \citep[e.g.,][]{Agertz:2009}.

While some of these physical mechanisms provide a compelling
explanation of the detected cool gas in hot halos, they remain
hypothetical due to a lack of empirical constraints.  To date, only a
handful of galaxies have been studied in detail to probe the origin of
their cool gas content.  For example, \cite{Nestor:2011} and
\cite{Gauthier:2013} considered a small sample of LRGs with associated
ultra-strong \MgII absorbers of $\ewr \apg 4$\,\AA, and found that
these LRGs preferentially reside in a group environment.  The presence
of a galaxy group is qualitatively consistent with the expectation of
the observed \MgII absorbers originating in stripped gas in the
intergroup medium.

As a first step toward a better understanding of the physical origin
of chemically-enriched cool gas in massive quiescent halos, we make
use of the vast spectroscopic data available in the public SDSS data
archive to obtain refined measurements of the incidence and covering
fraction of \MgII absorbing gas in LRG halos.  As described below, our
study is based on an unprecedentedly large sample of $\sim$ 38,000
LRGs spectroscopically identified at projected distances of $d< 500$ kpc
from the sightline of a background QSO.  Both LRGs and the background
QSOs are found in the SDSS spectroscopic catalog from DR12 (Alam
\etal\ 2015).  While the signal-to-noise ($S/N$) of the QSO spectra
varies according to the apparent brightnesses of the QSOs, we are able
to constrain the presence/absence of Mg\,II absorbers with rest-frame
absorption equivalent width $W_r(2796)\apg 0.3$ \AA\ for $\apg 35$\%
of the total LRG sample.
%  Our LRG-QSO pair sample is the largest to date, with
%more than two orders of magnitude increase in the sample size from
%previous studies \citep[e.g.][]{Bowen:2011,Gauthier:2011}. 
The unprecedentedly large LRG-QSO pair sample allows us to accurately
determine the mean incidence of extended \MgII gas around these
massive galaxies as a whole.  In addition, it enables a detailed study
of how the incidence and covering fraction of chemically-enriched cool
gas depend on the projected distance from central LRGs and on
additional galaxy properties, such as mass, emission-line properties,
and geometric alignments.  These observations provide important
insights into the origin of the observed chemically-enriched cool gas
in LRG halos.

%It is possible that LRGs associated with Mg II absorbers are within a
%subpopulation that contain ongoing star formation.

The paper is organized as follows.  In Section 2, we describe the
procedures to establish the close LRG-QSO pair sample, summarize the
general properties of the LRGs, and describe the absorption-line
measurements.  In Section 3, we inspect the photometric and spectral
properties of the LRGs and examine whether/how the observed \MgII
absorption strength and covering fraction in LRG halos are correlated
with galaxy properties.  Finally in Section \ref{section:discussion},
we consider different scenarios that can explain the presence of
chemically-enriched cool gas in the massive halos where these evolved
galaxies reside.  We adopt a standard $\Lambda$ cosmology,
$\Omega_M$ = 0.3 and $\Omega_\Lambda$=0.7 with a Hubble constant
$H_{\rm 0} = 70\rm \,km\,s^{-1}\,Mpc^{-1}$.

\begin{table*}
\centering
\caption{Definitions of Various Spectral Indices}
\label{table:index}
\centering {
\begin{tabular}{lrrrr}
\hline \hline
\multicolumn{1}{c}{Index} &  \multicolumn{1}{c}{Line passband (\AA)} & \multicolumn{1}{c}{Blue continuum (\AA)} & \multicolumn{1}{c}{Red continuum (\AA)} & \multicolumn{1}{c}{Reference} \\
%\multicolumn{1}{c}{(1)} & \multicolumn{1}{c}{(2)} & \multicolumn{1}{c}{(3)} & \multicolumn{1}{c}{(4)} & \multicolumn{1}{c}{(5)} \\
\hline 
\OII      & $3713.0-3741.0$ & $3653.0-3713.0$ & $3741.0-3801.0$ & \cite{Balogh:1999}\\
D4000  & \multicolumn{1}{c}{...} & $3850.0-3950.0$ & $4000.0-4100.0$ & \cite{Balogh:1999}\\
$\Hdelta_A$ & $4083.5-4122.3$ & $4041.6-4079.8$ & $4128.5-4161.0$ & \cite{Worthey:1997} \\
\OIII      & $4998.2-5018.2$ & $4978.2-4998.2$ & $5018.2-5038.2$ & \cite{Yan:2006}\\
\Halpha & $6554.6-6574.6$ & $6483.0-6513.0$ & $6623.0-6653.0$ & \cite{Yan:2006}\\
\NII    & $6575.3-6595.3$ & $6483.0-6513.0$ & $6623.0-6653.0$ & \cite{Yan:2006}\\
\hline
%		\hline \multicolumn{12}{l}{\bf Notes} \\
%		\multicolumn{12}{l}{$^\mathrm{a}$ Name of the COS quasar} \\
\end{tabular}
}
\end{table*}

\section{Data}
\label{section:sample}

We utilize existing spectroscopic data in the public SDSS archive to
characterize the CGM of massive galaxies.  Here we describe the
procedures that we followed for establishing the projected LRG and QSO pair
catalog and summarize the general properties of the LRGs in the pair
sample.  In addition, we describe the absorption-line measurements
that led to important constraints for the CGM around LRGs.

\subsection{The LRG-QSO Pair Catalog}
\label{section:LRGcat}
We first considered the galaxies and quasars from the Data Release
\citep[DR12,][]{{Alam:2015}} of SDSS, particularly those in the Baryon
Oscillation Spectroscopic Survey \citep[BOSS,][]{Dawson:2013}.  In
total, imaging and spectroscopic data were obtained in BOSS for $\sim$
1.5 million luminous galaxies at mean redshift
$\left<z\right>\approx0.6$.  The quasar sample includes about 150k
quasars from both SDSS-II and BOSS at $z<3.5$.  The new BOSS
multi-object spectrograph \citep[][]{Smee:2013} covers a wavelength
range from 3600\,\AA\ to 10400\,\AA, and enables observations of \MgII
absorbers at redshifts from as low as $z\approx\,$0.28 to $z\approx
2.7$.  We cross-matched spectroscopically identified galaxies with
background quasars to find projected pairs separated by $d<500$ kpc in
projected distance.  The background quasars and foreground galaxies
are drawn from the BOSS automated spectral classification and redshift
measurement pipeline (Bolton et al. 2012).  The maximum projected
distance of $d = 500$ kpc is chosen based on the expected size of a
typical LRG host dark matter halo.  We excluded galaxies that occur
within a line-of-sight velocity separation of $<10,000$ \kms\ from the
background QSO in order to exclude correlated QSO--galaxy pairs and to
avoid confusions between absorption features imprinted by the CGM of
LRGs and by QSO outflows.  The process yielded a total of 45757
galaxies at $d<500$ kpc from the sightline of a background QSO in the
SDSS sample.  Redshifts of the galaxies range from $z=0.30$ to
$z=1.42$.  Note that due to BOSS galaxy target selection, this galaxy
sample consists of $\approx 83\,\%$ LRGs and $\approx 17\,\%$ luminous
star-forming galaxies.

The spectroscopic targets of two primary BOSS galaxy samples, LOWZ
($z\apl\rm0.4$) and CMASS ($\rm 0.4\apl{\it z}\apl0.7$), were selected
using two sets of color-magnitude cuts similar to the LRG target
selection for SDSS-I/II \citep[][]{Eisenstein:2001}.  A crucial
difference is that the CMASS sample extends the SDSS-I/II LRG
selection to include blue objects.  As a result, while the majority of
targeted galaxies are LRGs, there is a non-negligible number of
massive star-forming galaxies that could potentially bias our results.
To identify LRGs from the initial BOSS galaxy sample, we further
applied a color selection criterion based on the intrinsic, rest-frame
$u-g$ color.  Using the Sbc galaxy template of \cite{Coleman:1980}, we
defined elliptical galaxies as those with rest-frame $u-g$ color
redder than the Sbc template, and star-forming galaxies as those with
bluer $u-g$ colors.  Specifically,
\begin{equation}
  \begin{split}
 &u-g > 1.18\ \ \ \rm \text{for\ elliptical/S0} \\
  &u-g \leq 1.18\ \ \ \rm \text{for\ disk/irregular/star-forming\ galaxies} \\
 \end{split}
\end{equation}
The rest-frame $u-g$ color of each galaxy was computed based on its
spectroscopic redshift from BOSS and interpolating between $u$-, $g$-,
$r$-, $i$-, and $z$-band composite model magnitudes from the SDSS
archive.  In addition to the rest-frame $u-g$ color selection, we also
restricted the sample selection to galaxies that do not exhibit strong
interstellar medium emission lines due to \OII\,$\lambda$3727,
\OIII\,$\lambda$5007, or H$\alpha$\,$\lambda\,6564$ at more than
5-$\sigma$ level of significance.  Some of these galaxies displayed
both dominant absorption features due to an evolved stellar population
and strong emission lines from H\,II regions.  The choice of a
5-$\sigma$ threshold is to ensure that we do not exclude more than 1\%
of evolved galaxy populations.  Details of measuring equivalent widths
of galaxy emission lines are presented in Section \ref{section:spec}
and definitions of various spectral indices are summarized in Table
\ref{table:index}.  Applying these additional cuts yielded a total of
38116 LRGs and $\apl 0.2$\% contamination by galaxies with younger
stellar populations based on visual inspections.  As described in the
next section (\S\ 2.2), 495 of these LRGs occur at a redshift where
the spectrum of the background QSO does not provide useful constraints
for the halo gas content.  Excluding these LRG-QSO pairs led to a
final sample of 37621 LRGs for the subsequent CGM studies.

\begin{figure*}
	\centering
	\includegraphics[scale=0.46]{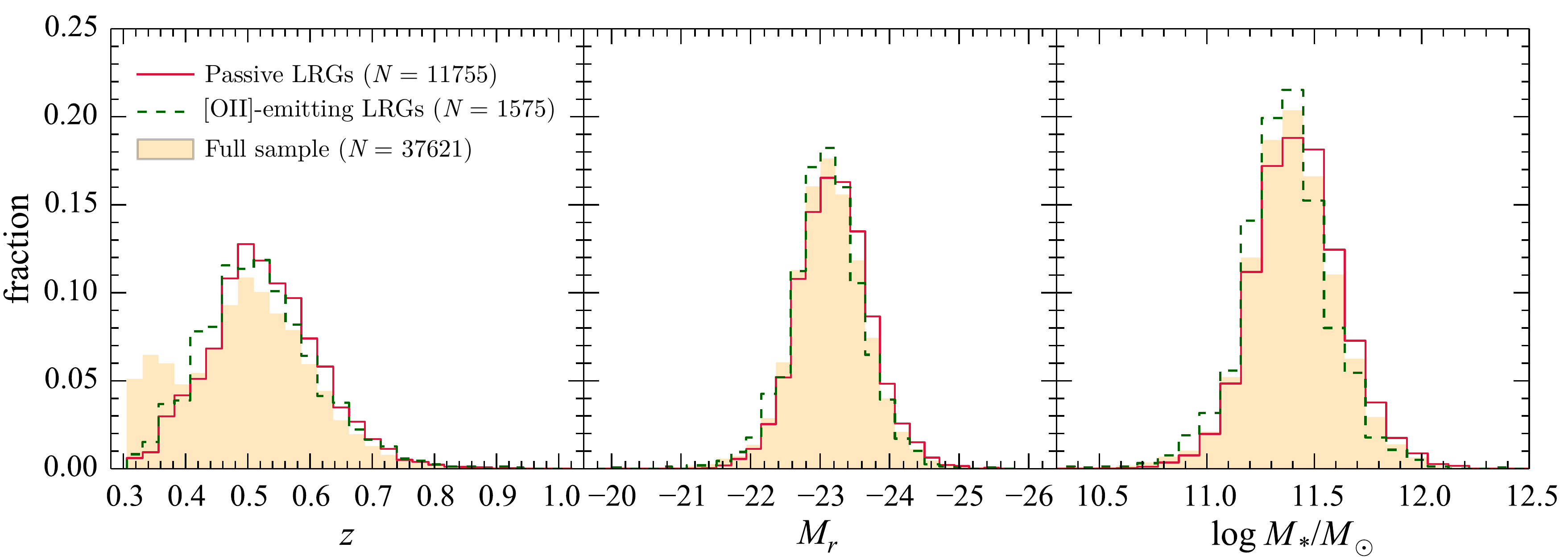}
	\caption{Distributions in redshift $z$, $r$-band absolute
          magnitude $M_r$, and stellar mass $M_*$ of LRGs in our
          study.  The full sample consists of 37621 LRGs from a
          combination of SDSS LOWZ sample that dominates the LRG
          population at $z\apl 0.4$ and the CMASS sample that
          dominates at $z=0.4-0.7$ (left panel).  Considering only
          LRGs with sufficiently high $S/N$ background QSO spectra for
          detecting Mg\,II absorbers as weak as $W_0=0.3$ \AA\ yields
          a total sample of 13330 LRGs, 1575 of which are
          \OII-emitting (open, dashed histograms) and 11755 are
          passive LRGs with no trace of \OII\ emission at more than a
          2-$\sigma$ level of significance (open, red histograms).
          The redshift distributions of these LRGs show that the
          $W_0=0.3$ \AA\ cut removes a significant fraction of LRGs
          from the LOWZ sample, due to the rapidly declining
          throughput at $\lambda\apl 4000$ \AA\ of the SDSS
          spectrograph (Smee \etal\ 2013) that prohibited us from
          detecting relatively weak Mg\,II absorption features at
          $z\apl 0.4$.  All three LRG samples share similar
          distributions in $M_r$ (middle panel) and $M_*$ (right
          panel).}
	\label{figure:statistics}
\end{figure*}

To characterize the general properties of these LRGs, we computed the
rest-frame absolute $r$-band magnitude $M_r$ and stellar mass $M_*$
based on the observed SDSS $g$-, $r$-, $i$-, and $z$-band magnitudes,
using the {\it K}-correct code \citep[][]{Blanton:2007}.  We excluded
the $u$-band from the photometric analysis, because $\approx 99$\% of
the LRGs are not detected at $\apg\,5\,\sigma$ level of significance
in the $u$-band.  Distributions in redshift, $M_r$, and $M_*$ for the
full sample are presented in Figure \ref{figure:statistics} in filled
histograms.  The redshift distribution of our LRGs (left panel of
Figure 1) clearly shows a double-peak feature, with the majority
selected from the CMASS sample at $z=0.4-0.7$ and some fraction from
the SDSS LOWZ sample at $z\apl 0.4$.  We note, however, that a
substantial fraction of LRGs from the LOWZ program do not have
sensitive constraints for their associated Mg\,II absorption features,
because at this redshift range the Mg\,II doublet transitions occur at
$\lambda<4000$ \AA, where the throughput of the spectrograph declines
rapidly (Smee \etal\ 2013).  Stellar masses of the full LRG sample
span a range from $\log\,M_*/M_\odot<11$ to $\log\,M_*/M_\odot\approx
12$ (right panel of Figure 1) with a mean of
$\langle\,\log\,M_*/M_\odot\,\rangle=11.4$ and a dispersion of 0.2
dex.

While LRGs exhibit spectral features that are typical of an old and
passively evolved stellar population, roughly 10\% of the LRG
population also exhibit emission lines from recent star formation or
AGN activity (e.g.\ Roseboom \etal\ 2006).  To isolate quiescent
galaxies with little/no on-going star formation, we further divided
the final LRG sample into two subsamples based on the significance of
the observed [O\,II] emission.  The procedure yielded 4994 LRGs with
\OII\ detected at greater than 2-$\sigma$ level of significance, and
32627 passive LRGs without detected \OII.  
%Comparing the passive and
%\OII-emitting LRG samples shows that galaxies in the two samples share
%the same distribution functions in redshift, $M_r$, and $M_*$ (Figure
%1).

In subsequent discussions, we focus our analysis on a subsample of
LRGs for which sensitive constraints for their halo gas content can be
placed using the spectrum of a background QSO (more details are
described in Section \S\ 2.2).  Considering only LRGs with
sufficiently high $S/N$ background QSO spectra for detecting Mg\,II
absorbers as weak as $W_0=0.3$ \AA\ led to a total sample of 13330
LRGs (Table 2).  Figure 1 demonstrates that for LRGs with sensitive
absorption-line constraints, both passive (open, red histograms) and
[OII]-emitting (open, dashed histograms) LRGs exhibit similar
distributions in redshift, $M_r$, and $M_*$.  Specifically,
\OII-emitting LRGs span a range in redshift from $z=0.31$ to $z=0.94$
with a median of $\langle\,z\,\rangle_{\rm med}= 0.51$, while passive
LRGs cover a redshift range from $z=0.31$ to $z=0.99$ with
$\langle\,z\,\rangle_{\rm med}= 0.52$.  In terms of stellar mass,
\OII-emitting LRGs are well described by a Gaussian distribution with
a mean of $\langle\,\log\,(M_*/M_\odot)\,\rangle = 11.37$ and a
dispersion of 0.18 dex while passive LRGs without detected \OII\ are
well described by a Gaussian distribution of
$\langle\,\log\,(M_*/M_\odot)\,\rangle=11.42$ and a dispersion of 0.2
dex (Table 2).  The difference in the mean stellar masses is
negligible, given the $\approx$ 0.2 dex scatter introduced in
estimating photometric stellar masses \citep[][]{Blanton:2007}.

Finally, we note that the previous 5-$\sigma$ equivalent width cut had
excluded $\sim 1$\% of strong \OII-emitting LRGs with underlying old
stellar populations.  Including these strong \OII-emitting LRGs will
not significantly alter the statistical properties of \OII-emitting
LRGs in our sample, but will likely increase the differences found in
the CGM between \OII-emitting and non-\OII emitting LRGs.  We defer
the analysis of those strong \OII-emitting LRGs to a future paper.

\begin{table*}
\centering
\caption{Summary of the LRG Samples}
\label{table:LRGsample}
\centering \resizebox{6.in}{!}{
\begin{tabular}{ccrrrcc}
\hline \hline
{$W_0$ (\AA)} & LRG Type & \multicolumn{1}{c}{{\it N}(total)}  & \multicolumn{1}{c}{{\it N}(upper limit)} & \multicolumn{1}{c}{{\it N}(detection)} & $\rm \langle log\,M_*/M_\odot \rangle$ & $\rm \sigma(\langle log\,M_*/M_\odot \rangle) $\\
%(1) & (2) & \multicolumn{1}{c}{(3)} & \multicolumn{1}{c}{(4)} & \multicolumn{1}{c}{(5)} & (6) & (7) \\
\hline 
Any & Full & 37621 & 36911 & 710 & 11.40 & 0.19 \\
\hline 
0.3 & \OII-emitting & 1575 & 1499 & 76 & 11.37 & 0.18\\
0.3 & Passive & 11755 & 11215 & 540 & 11.42 & 0.20\\
0.3 & All & 13330 & 12714 & 616 & 11.42 & 0.20\\

%\hline
%0.1 & All & 1065 & 166 & 1231 & 100\\
%& \OII\ emitting & 131 & 20 & 151 & 12.3\\
%& passive & 934 & 146 & 1080 & 87.7\\
\hline
%		\hline \multicolumn{12}{l}{\bf Notes} \\
%		\multicolumn{12}{l}{$^\mathrm{a}$ Name of the COS quasar} \\
\end{tabular}
}
\end{table*}

\subsection{Extended \MgII Halos around LRGs}% Absorber Pair Sample}
\label{section:MgII}

To constrain the halo gas content of LRGs, we took each projected
LRG--QSO pair from \S\ 2.1 and manually searched for the corresponding
\MgII absorption features at the redshift of the LRG in the spectrum
of the background QSO.  Our search window covers a radial velocity
interval of $\Delta\,v = \pm 1000$ \kms, centered at the systemic
redshift of the LRG.  This large search window was chosen to include
the vast majority of \MgII absorption systems originating in LRG halos.
It is based on the expected velocity dispersion of $\sigma$ $\sim$ 350
\kms\ for virialized gas in halos of $\sim 10^{13} {\rm M_\odot}$.
%Expanding the search window to $1000 < |\Delta\,v| \leq 2000$ \kms\
%uncovered no more than 10 \MgII-absorbing LRGs.  
In addition, as described in \S\ 3.2 below, the detected \MgII
absorbers exhibit a simple Gaussian distribution in velocity offset
from the systemic redshifts of the LRGs, which is characterized by a
velocity dispersion of $\sigma_v\approx 165$ \kms\ (see Figure 3
below).  We are therefore confident that the adopted search window was
sufficiently large to find associated absorbers.

When a \MgII absorber was found in the QSO spectrum, we measured the
rest-frame absorption equivalent width of the 2796 $\rm \AA$ member,
$W_r(2796)$, and determined the absorber redshift based on the
best-fit line centroid of a Gaussian profile.  For 20 LRGs, multiple
\MgII absorbers were found within the search window.  We adopt the
velocity centroid of the strongest component as the systemic velocity
of the absorbing gas.  When no \MgII features were detected, we record
a 2-$\sigma$ upper limit of the underlying absorber strength over a
wavelength window defined by the width of a spectral resolution
element of SDSS spectra (${\rm FWHM}=150$ \kms), based on the
associated 1-$\sigma$ error spectrum.

In the full LRG sample, two groups of LRGs were found to have
corresponding \MgII absorbers within $\Delta\,v<1000$ \kms.  For these
two cases, we assigned absorbers to the LRGs at the smallest projected
distance and excluded the remaing LRGs from the sample.  The procedure
identified 710 \MgII absorbers and 36911 upper limits in the
vicinities of 37621 LRGs.  We were unable to obtain significant
constraints for \MgII absorbers around 478 LRGs, where the
accompanying QSO spectra have extraction defects, broad absorption
line complexes, or are contaminated by other absorption transitions
(such as \CIV\,$\lambda\lambda$\,1548,1550) from a different
redshift.  For 17 LRGs, the corresponding \MgII absorption was
expected to fall close to a prominent QSO emission line.  Because of
significant uncertainties in the continuum near the peak of a QSO
emission line, these LRGs are also excluded from the analysis.  

The constraints we were able to place for the presence or absence of
extended Mg\,II absorbing gas around the LRGs were also non-uniform.
Because of varying qualities (in terms of $S/N$) of the absorption
spectra at different wavelengths and between different QSOs, there
exists a substantial scatter in the upper limits we were able to place
for those 36911 LRGs.  For a large fraction of the LRGs ($\approx 2/3$
of the sample), the apparent, low $S/N$ spectra of the QSOs prohibited
us from placing sensitive constraints for the underlying Mg\,II
absorbers.

To facilitate a uniform analysis, we focused our subsequent studies on
a homogeneous sample of LRGs with sufficiently high $S/N$ background
QSO spectra that allow a minimum detection threshold in rest-frame
absorption equivalent width of $W_0=0.3$ \AA\ for the Mg\,II
absorption features.  A summary of the LRG samples and the Mg\,II
survey result is shown in Table \ref{table:LRGsample}.  Under the
$W_0=0.3$ \AA\ minimum quality cut, we have a sample of 11755 passive
LRGs without \OII\ detected at a more than 2-$\sigma$ level and 1575
\OII-emitting LRGs.  The fraction of \OII-emitting LRGs ($\approx
13$\%) is consistent with what was found in the 2dF$-$SDSS LRG and QSO
Survey \citep[][]{Roseboom:2006}.

To compare the general properties of the \OII-emitting and passive LRG
samples, we include their distributions in redshift, $M_r$ and $M_*$
in Figure 1 and Table 2.  The passive LRGs with sensitive Mg\,II
absorption constraints are shown in open, red histograms, and the
\OII-emitting LRGs are shown in dashed, green histograms.  The
comparisons confirm that with the additional $W_0=0.3$ \AA\ selection
criterion, the underlying redshift and $M_*$ distributions of the
resulting \OII-emitting and passive LRG subsamples remain the same.

%the Mstar is compared to multiple groups and all yield similar narrow
%result .with overall offset.  But doesn't ulter the result.
% talk about Mvir, consistent with xxx.

\section{Analysis}
\label{section:result}

The procedures described in Section \ref{section:sample} established a
sample of 13330 LRGs with sensitive background QSO spectra available
for constraining the presence/absence of \MgII absorbers of
$\ewr\,\geq 0.3$ \AA.  Of these, 1575 show \OII\ emission in the LRG
spectra and 11755 LRGs appear to contain a passive and old stellar
population with no \OII-emission detected at greater than a 2-$\sigma$
level.  As summarized in Table 2, the passive and \OII-emitting LRGs
share a very similar distribution in stellar mass with a mean of
$\langle\,\log\,M_*/M_\odot\,\rangle=11.4$ and a dispersion of 0.2 dex
(see also the right panel of Figure 1).  In addition, our search of
Mg\,II absorption features in the vicinities of these LRGs yielded 76
detections around \OII-emitting LRGs (a rate of incidence $\approx
4.8$\%) and 540 detections around passive LRGs ($\approx 4.6$\%).

In this section, we first inspect the general properties of passive
and \OII-emitting LRG samples, focusing primarily on their respective
stellar populations and star formation histories.  Then we examine
Mg\,II absorption properties in LRG halos and investigate possible
correlations between LRG properties and the observed \MgII absorption
properties in LRG halos.

%In section \ref{section:spec}, we first perform stacks of LRG spectra
%and determine their spectral properties.  We show that LRGs are indeed
%dominated by old stellar populations.  In Section \ref{section:CGM},
%we characterize the CGM properties in massive LRG host halos.  We
%present the rest-frame equivalent width of \MgII absorbers \ewr\ as a
%function of projected distance $d$ in Figure \ref{figure:EWrho03}, the
%relative velocity dispersion of \MgII absorbers in Figure
%\ref{figure:vdiff} and the covering fraction versus $d$ in Figure
%\ref{figure:kappa03}.  Finally, we explore the azimuthal dependence of
%the \MgII gas covering fraction in various distances in Section
%\ref{section:azimuthal}.  We show that these much larger samples allow
%us to have unprecedented constraints on the incidence of \MgII
%absorbers in massive halos and, for the first time, to reveal the
%correlation between the presence/absence of \MgII absorbing gas in LRG
%host halos and \OII\ emission features in LRGs.
 
 \begin{figure*}
	 \includegraphics[angle=0,scale=0.7]{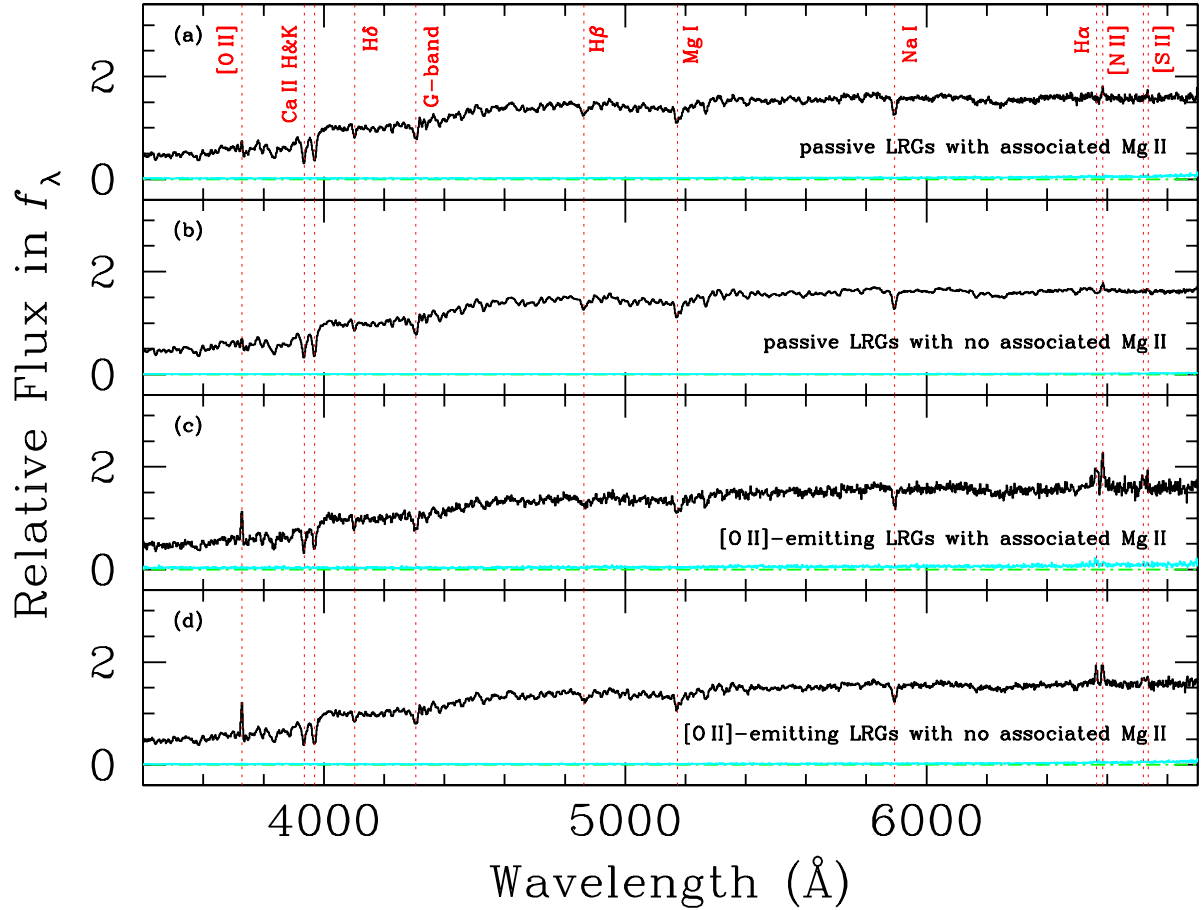} 
         \caption{Stacked rest-frame spectra of LRGs at $d<200$ kpc
           from a background QSO sightline with sufficiently high
           $S/N$ absorption spectra available for detecting a \MgII
           absorber of $W_r(2796)>0.3$ \AA.  The corresponding
           1-$\sigma$ dispersion in each stack is shown in cyan at the
           bottom of each panel.  Panel (a) shows the median stack of
           204 passive LRGs with associated Mg\,II absorbers, and
           panel (b) displays the median stack of 1567 passive LRGs
           without associated \MgII absorbers.  The stacked spectra
           are characterized by prominent absorption features due to
           Ca\,{\scriptsize II} H $\&$ K, G-band, Mg\,{\scriptsize I}
           and Na\,{\scriptsize I} that indicate a predominantly old
           stellar population, as well as a relatively weak Balmer
           absorption series.  Neither \OII\ nor H$\alpha$ emission
           line is detected in the stacks but we note the presence of
           [N\,II]\,$\lambda\,6585$ emission.  Panels (c) and (d)
           display the stacked spectra of \OII-emitting LRGs with and
           without associated \MgII absorbers, respectively.  A total
           of 41 LRG spectra are included in the stack presented in
           panel (c) and 240 LRGs in panel (d).  Similar to passive
           LRGs, the stacked spectra of \OII-emitting LRGs display
           prominent absorption features that indicate the presence of
           evolved stellar populations, but at the same time modest
           emission features due to H$\alpha$, [N\,II], and [S\,II]
           are also detected.  The observed relatively weak H$\alpha$
           emission together with a prominent [N\,II]\,$\lambda$\,6585
           emission feature suggests a subdominant presence of
           LINERs. }
	\label{figure:spec_med}
\end{figure*}

 \subsection{Mean Stellar Populations of the LRGs}
\label{section:spec}
%In the previous section, we found the
%covering fraction of both LRG samples flattens out
%at distances larger than $\approx\,200\, \rm kpc$,
%where the contribution from background \MgII absorbers becomes significant.
While LRGs are generally understood as high-redshift counterparts of
nearby massive elliptical galaxies that are made of predominantly old
stellar populations with little/no star formation, the visual
inspectation described in \S\ 2.1 has confirmed previous findings that
roughly 10\% of these LRGs exhibit \OII\ emission features that
suggest a modest amount of star formation among otherwise evolved
stellar populations (e.g.\ Roseboom \etal\ 2006).  Because a primary
goal of targeting gaseous halos around LRGs is to investigate whether
and how chemically-enriched absorbing gas can arise at large distances
from evolved galaxies in the absence of active on-going star
formation, it is necessary to first characterize the mean star
formation history and mean stellar population of \OII-emitting and
passive LRG samples before discussing the properties of their halo gas
content.

%To compare the stellar populations of LRGs associated
%with \MgII absorption systems at $d<200\, {\rm kpc}$ and without in both \OII\ emitting
%LRGs and passive LRGs, 

To examine the stellar population, we form a stacked spectrum of each
of the LRG subsamples and measure its spectral properties.  To
generate a stacked spectrum, we first mask out strong sky emission
lines. %NEED TO DO IT, ASK!
We then shift the observed spectrum to the rest frame of the galaxy
and adopt a constant pixel resolution of $\Delta\,\lambda=2$ \AA,
corresponding to roughly $\Delta\,v=150$ \kms\ at 4000 \AA, and $75$
\kms\ at 8000 \AA.  Individual LRG spectra were normalized using the
mean flux over $\lambda=4100-4300$ \AA, and median combined to form
the final stacked spectrum.

For each stacked spectrum, we measure the D4000 index and the
equivalent widths of \OII, $\Hdelta_A$, \Halpha, and \NII\ lines.  For
each emission line, the continuum is determined by measuring the mean
flux level in two sidebands and interpolating across the central
passband.  Then we sum over the continuum-normalized flux in the
central passband to obtain the equivalent width of the line.  For
\OII\ and \Hdelta, we use the standard definitions from
\cite{Balogh:1999} and \cite{Worthey:1997}, respectively.  For
\Halpha\ and \NII, we adopt the window definitions from
\cite{Yan:2006}.  We define the D4000 index using the narrow
definition introduced by \cite{Balogh:1999}, namely the ratio of the
fluxes in two 100-$\rm \AA$ windows centered at 4050 and 3900 $\rm
\AA$.  The definitions of the passbands and sidebands are summarized
in Table \ref{table:index}.

\begin{table*}
\centering
\caption{Summary of LRG properties from stacked spectra}
\label{table:spec}
%\centering \resizebox{6.9in}{!}{
\centering \resizebox{6in}{!}{
\begin{tabular}{ccccccccc}
\hline \hline
 &  & &  & EW([O {\scriptsize II}]) & EW(H$\delta_A$)   & EW(\Halpha)$^a$ & EW(\NII) & ${\rm SFR}_\Halpha^b$ \\
Spectral Type & \MgII Absorption & & D4000 & (\AA) & (\AA) & (\AA) & (\AA) & (${\rm M}_{\odot}\,/\,{\rm yr}$) \\
(1) & (2) & & (3) & (4) & (5) & (6) & (7) & (8) \\
\hline 
Passive LRGs  & detection & & 1.63 $\pm$ 0.01 & 0.61 $\pm$ 0.28 & 1.44 $\pm$ 0.13 & $-1.46\pm 0.26$  & $-1.18\pm 0.19$ & $0.48\pm 0.09$ \\
Passive LRGs  & non-detection & & 1.64 $\pm$ 0.01 & 0.78 $\pm$ 0.10 & 1.05 $\pm$ 0.04 & $-1.03\pm 0.08$  & $-0.80\pm 0.06$ & $0.34\pm 0.03$ \\
\OII-emitting LRGs  & detection & & 1.53 $\pm$ 0.03 & $-5.45 \pm 0.53$ & 2.02 $\pm$ 0.39 & $-4.24 \pm 0.94$  & $-4.19\pm 0.48$ & $1.40\pm 0.31$\\
\OII-emitting LRGs  & non-detection & & 1.52 $\pm$ 0.01 & $-4.98 \pm 0.24$ & 1.77 $\pm$ 0.15 & $-3.28\pm 0.33$  & $-2.70\pm 0.19$ & $1.08\pm 0.11$ \\
% &  & &  & EW([O {\scriptsize II}]) & EW(H$\rm_\delta$)   & EW(\Halpha) & SFR$\rm _{[O\,{\scriptsize II}]}$ \\
%Spectral Type & \MgII Absorption & & D4000 & (\AA) & (\AA) & (\AA) & ($\rm M_{\odot}\, yr^{-1}$) \\
%(1) & (2) & & (3) & (4) & (5) & (6) & (7) \\
%\hline
%Passive LRGs  & detection & & 1.63 $\pm$ 0.01 & 0.61 $\pm$ 0.28 & 1.44 $\pm$ 0.13 & 0.09 $\pm$ 0.26  & $<$ 0.08 \\
%Passive LRGs  & non-detection & & 1.64 $\pm$ 0.01 & 0.78 $\pm$ 0.10 & 1.05 $\pm$ 0.04 & 0.26 $\pm$ 0.08  & $<$ 0.03 \\
%\OII-emitting LRGs  & detection & & 1.53 $\pm$ 0.03 & -5.45 $\pm$ 0.53 & 2.02 $\pm$ 0.39 & -3.27 $\pm$ 0.94  & 0.77 $\pm$ 0.23\\
%\OII-emitting LRGs  & non-detection & & 1.52 $\pm$ 0.01 & -4.98 $\pm$ 0.24 & 1.77 $\pm$ 0.15 & -2.09 $\pm$ 0.33  & 0.70 $\pm$ 0.20 \\
\hline
%		\hline \multicolumn{12}{l}{\bf Notes} \\
%		\multicolumn{12}{l}{$^\mathrm{a}$ Name of the COS quasar} \\
\multicolumn{9}{l}{$^a$Corrected for stellar absorption based on a fit to the higher-order Balmer absorption series.} \\
\multicolumn{9}{l}{$^b$Based on the assumption that the observed \Halpha\ emission traces young stars and applying the scaling relation of Kennicutt \& Evans (2012).  But}  \\% the observed} \\
\multicolumn{9}{l}{\ \ because the observed emission-line ratios resemble LINER-like galaxies (e.g., Yan \etal\ 2006), the inferred SFR represent only an upper limit.}\\
\end{tabular}
}
\end{table*}

To estimate uncertainties in the measured line indices, we perform a
bootstrap analysis and repeat the stacking procedure 1000 times.  We
record the 1-$\sigma$ dispersions in the measured spectral indicies as
the measurement uncertainties.  The results are summarized in Table
\ref{table:spec}.  We have also experimented with forming a mean
(rather than median) stack and found that the measurements remain
consistent within uncertainties.  Note that throughout this paper we
define a negative equivalent width as emission and a positive value as
absorption.
%in order to avoid boosting the emission line
%and depressing the absorption line in the common S/N weighting scheme.
% do we want pixel by pixel weighting?

The resulting median stacks of LRG spectra are presented in Figure
\ref{figure:spec_med}, including passive LRGs with associated \MgII
absorbers in panel (a), passive LRGs without associated \MgII
absorbers in panel (b), \OII-emitting LRGs with detected \MgII
absorbing gas in panel (c), and \OII-emitting LRGs without associated
\MgII absorbers in panel (d).  The stacked spectra presented in Figure
\ref{figure:spec_med} include only LRGs that occur at $d<200$ kpc from
a background QSO sightline with sufficiently high $S/N$ absorption
spectra available for detecting a MgII absorber of $\ewr>0.3$ \AA.
This is motivated by the apparent difference in the observed covering
fraction of \MgII absorbers at $d<200$ kpc around different LRG
samples, as shown in the following section (\S\ 3.2).  The goal is to
examine whether the observed difference in extended \MgII absorbing
gas is related to the stellar population in the galaxies.

%While the median combined spectra have higher S/N, the mean combined spectra tend
%to have more prominent emission line features as expected.
%Regardless of the methods used, we obtain 
%consistent measurements of spectral properties,which would not change the discussion
%in this section.
Figure \ref{figure:spec_med} shows that all the LRG samples exhibit
prominent absorption features due to Ca\,{\scriptsize II} H $\&$ K, G
band, Mg\,{\scriptsize I} and Na\,{\scriptsize I}, indicating that the
spectra are dominated by old stellar populations.  We make use of the
$\Hdelta_A$ and D4000 spectral indicators as diagnostics of recent
star formation histories.  The D4000 index is known to be sensitive to
recent star formation, with lower values indicating an increasing
presence of a young stellar population.  The \Hdelta\ absorption line,
on the other hand, occurs in galaxies that have experienced a burst of
star formation $\sim 0.1-1\,\rm Gyr$ ago.  The absorption strength is
expected to peak at $\sim 1\,\rm Gyr$ when hot O and B stars have left
the main sequence, and decline with increasing age afterward.  The
\OII-emitting LRGs have D4000\,$\approx$\,1.5 and \Hdelta\ equivalent
width of EW($\Hdelta_A$)\,$\approx2$, whereas passive LRGs have
D4000\,$\approx$\,1.6 and EW($\Hdelta_A$)\,$\approx1.2$.  No
significant difference is found for the D4000 index and only marginal
difference is seen in \Hdelta\ between \MgII-absorbing and
non-absorbing passive LRGs.

Following the diagnostics described in \cite{Kauffmann:2003}, we
estimate a mean stellar age of $\apg\,1\,\rm Gyr$ based on the mean
spectral indices observed in the stacked LRG spectra.  The estimated
mean stellar age is consistent with the conclusions of
\cite{Gauthier:2011} for 37 individual LRGs based on a stellar
population synthesis analysis.  If the \OII-emitting and passive LRG
samples share a similar metallicity, then the smaller mean D4000
indices and higher EW(\Hdelta) in \OII-emitting LRGs suggest an on
average younger stellar population in these galaxies than in passive
LRGs.
%Note that one caveat of the measurement is that for older
%stellar populations, the indices depend strongly on metallicity \citep[e.g.][]{Kauffmann:2003}.

We also investigate the emission line properties of different LRG
samples.  For passive LRGs, we do not uncover \OII\ emission even in
the high $S/N$ stacked spectra.  For \OII-emitting LRGs, we uncover 
relatively weak \Halpha\ emission in the stacked spectra.  Both
passive and \OII-emitting LRGs exhibit a modest
[N\,II]\,$\lambda\,6585$ emission feature.  
%The relatively weak
%\Halpha\ emission suggests a subdominant presence of recent star
%formation or active galactic nuclei (AGN).  
After correcting for stellar absorption using the observed H$\beta$,
H$\gamma$, and H$\delta$ absorption features, we recover the
underlying H$\alpha$ emission flux for all LRG subsamples.  The
stellar absorption-corrected H$\alpha$ emission equivalent width for
each subsample is presented in column (6) of Table 3.  Assuming that
the observed \Halpha\ emission traces on-going star-formation in the
LRGs and applying the scaling relation from Kennicutt \& Evans (2012),
we infer an unobscured star formation rate (SFR) based on the observed
EW(\Halpha) and mean $M_r$.
%We note that the \Halpha\ line is only
%observed in about half of the LRGs that are located at $z<0.5$ and
%that many of the observed lines occur in noisy regions of the SDSS
%spectra due to poor sky subtraction.  Therefore, we do not consider an
%\Halpha-based SFR measurement a robust one in this case.  
The estimated mean ${\rm SFR}_\Halpha$ of each LRG sample is presented
in Table \ref{table:spec}.  We find that \OII-emitting LRGs have a
mean SFR as high as ${\rm SFR}_\Halpha\approx 1-1.5\, {\rm
  M}_\odot\,{\rm yr}^{-1}$ and passive LRGs have a mean SFR as high as ${\rm
  SFR}_\Halpha \approx 0.3-0.4\, {\rm M}_\odot\,{\rm yr}^{-1}$.

%The modest SFR derived for passive and \OII-emitting LRGs is consistent with the
%relatively younger age inferred for the \OII-emitting LRGs based on
%absorption-line indices.  However, among the \OII-emitting LRG
%samples, we do not find an obvious correlation between the
%presence/absence of \MgII absorbing halo gas and the on-going SFR.
%But because the SDSS fibers cover only a sky area of $2''$ in diameter
%($\approx$ 12 kpc in diameter at $z\approx0.5$), it is possible that
%the SDSS spectra have missed star formation that occurs in the
%outskirts of LRGs.  We consider this an unlikely scenario.  In the
%recent analysis, \cite{Tal:2011} used stacked images of more than
%42000 LRGs and detected excess light out to 100 kpc of the LRGs.
%While it is possible that the lack of dependence is due to the missing
%light in the outskirts, \cite{Tal:2011} found the colors of the
%extended light are consistent with the colors of the LRGs, suggesting
%the stellar population does not vary significantly with radius in the
%LRG halos.

However, many local elliptical galaxies and passive red galaxies at
higher redshifts display emission features that resemble the
low-ionization nuclear emission-line regions (LINERs; e.g., Sarzi
\etal\ 2006; Yan \etal\ 2006).  Searches for radio emission based on
stacks of FIRST images of the LRGs have also continued to uncover
faint radio fluxes in these galaxies (e.g., Hodge \etal\ 2008, 2009).
The observed high [N\,II]\,/\,\Halpha\ ratio, together with a low
[O\,III]\,/\,[O\,II] ratio, in our stacked spectra of \OII-emitting
galaxies (bottom two panels of Figure \ref{figure:spec_med}) indeed
confirms previous findings that these LRGs are LINER-like galaxies
(e.g.\ Johnston \etal\ 2008; Hodge \etal\ 2008).
%At the same time, the
%observed low [O\,III]\,/\,[O\,II] ratios resemble the spectra of
%LINER-like galaxies, rather than star-forming regions or Seyfert
%galaxies (e.g., Yan \etal\ 2006).  It is therefore reasonable to
%expect that the LRG halos are being regulated by AGN winds that
%provide an additional heating source.
Therefore, the observed \OII\ and \Halpha\ emission features are most
likely due to underlying active galactic nuclei (AGN) or LINERs, and the
SFR estimated based on the observed \Halpha\ emission flux only
represents an upper limit.

%  is seen in many elliptical galaxies (LINER-like): Yan+05,12
% talk about the origin of OII
% does [O ii] also indicate star formation in these galaxies? It has long been realized that star formation is not the only possible source of [O ii] emission in galaxies
% Active galactic nuclei (AGNs, especially low-ionization nuclear emission- line regions [LINERs]), fast shock waves, postYasymptotic giant branch ( post-AGB) stars, and cooling flows might also produce [O ii] emission. (Yan06)
%But Salder et al (2006) shows that the emission
%LRGs sample does not show a higher fraction of radio emission
%compared to the overall sample.
%Although we will be confident that
%the ones with radio emission are due to
%AGN activity,
%the converse, that is, lack of radio emission constitutes
%a lack of AGN activity is not true.
%Alternatively, we can check the correlation between
%[OII] emission and lower D4000 indices.  Unless we
%are observing an LRG during the very short period 
%of time at the very beginning of a star formation episode
%before substantial numbers of young stars have been formed,
%there should be a clear relation between the D4000 index 
%and current star formation activities.  However, if [OII] emission
%originates from AGN activities, there would be no relation.\\ 
%

\begin{figure*}
	 \includegraphics[angle=0,scale=0.5]{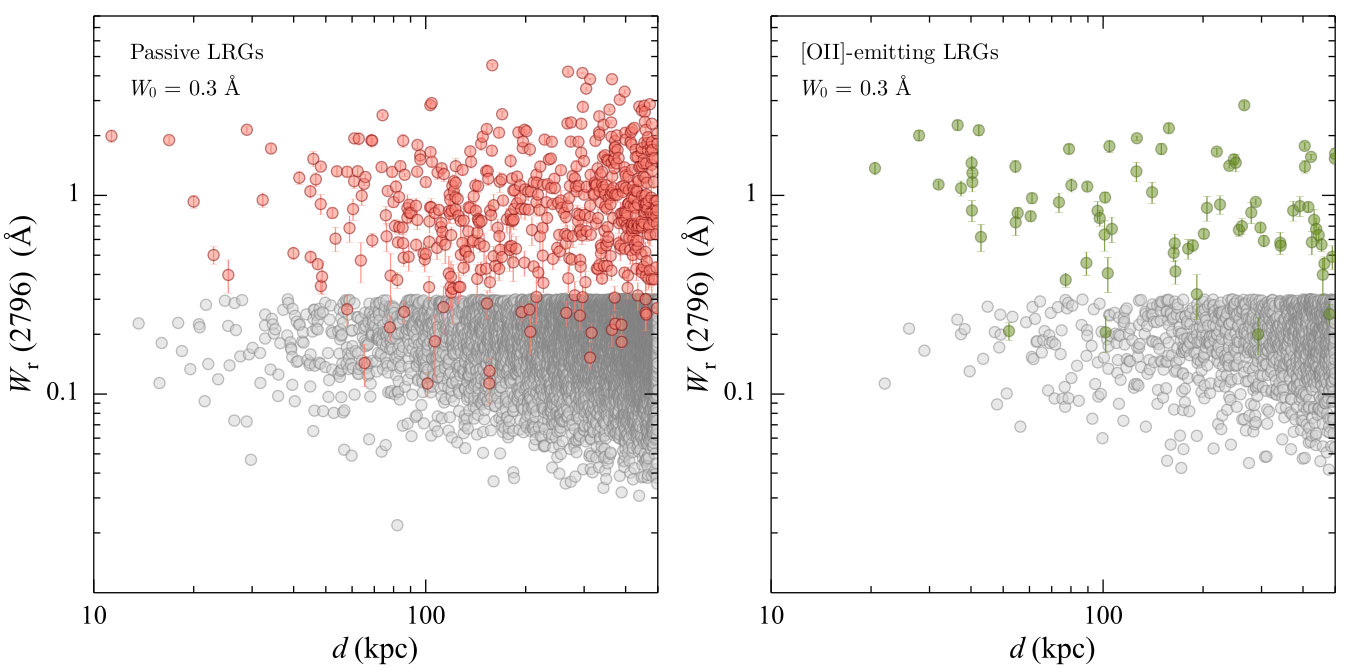} 
         %\ \ \ \ \  \includegraphics[angle=0,scale=0.46, trim=0in 0.0in 0in 0in]{EWrhoOII03.pdf}
         \caption{Rest-frame absorption equivalent width $\ewr$ versus
           projected distance $d$ for LRGs with sufficiently high
           $S/N$ background QSO spectra for detecting Mg\,II absorbers
           as weak as $W_0=0.3$ \AA.  Absorption observations for
           passive LRGs are shown in the {\it left} panel.
           Observations of \OII-emitting LRGs are shown in the {\it
             right} panel.  LRGs with detected Mg\,II absorbers are
           shown in filled circles with errorbars representing
           measurement uncertainties.  For non-detections, we use
           light grey circles to indicate the 2-$\sigma$ upper limits.
           Note that a significant fraction of the original SDSS LRG
           sample have poor-quality QSO spectra that result in upper
           limits exceeding 0.3 \AA.  These LRG--QSO pairs offer
           little/no constraints for the underlying absorber strengths
           in LRG halos and are therefore excluded from the plots for
           clarity. }
	\label{figure:EWrho03}
\end{figure*}

\subsection{Properties of \MgII Absorbing Gas  in LRG Halos}
\label{section:CGM}

\begin{figure*}
\centering
\includegraphics[scale=0.5]{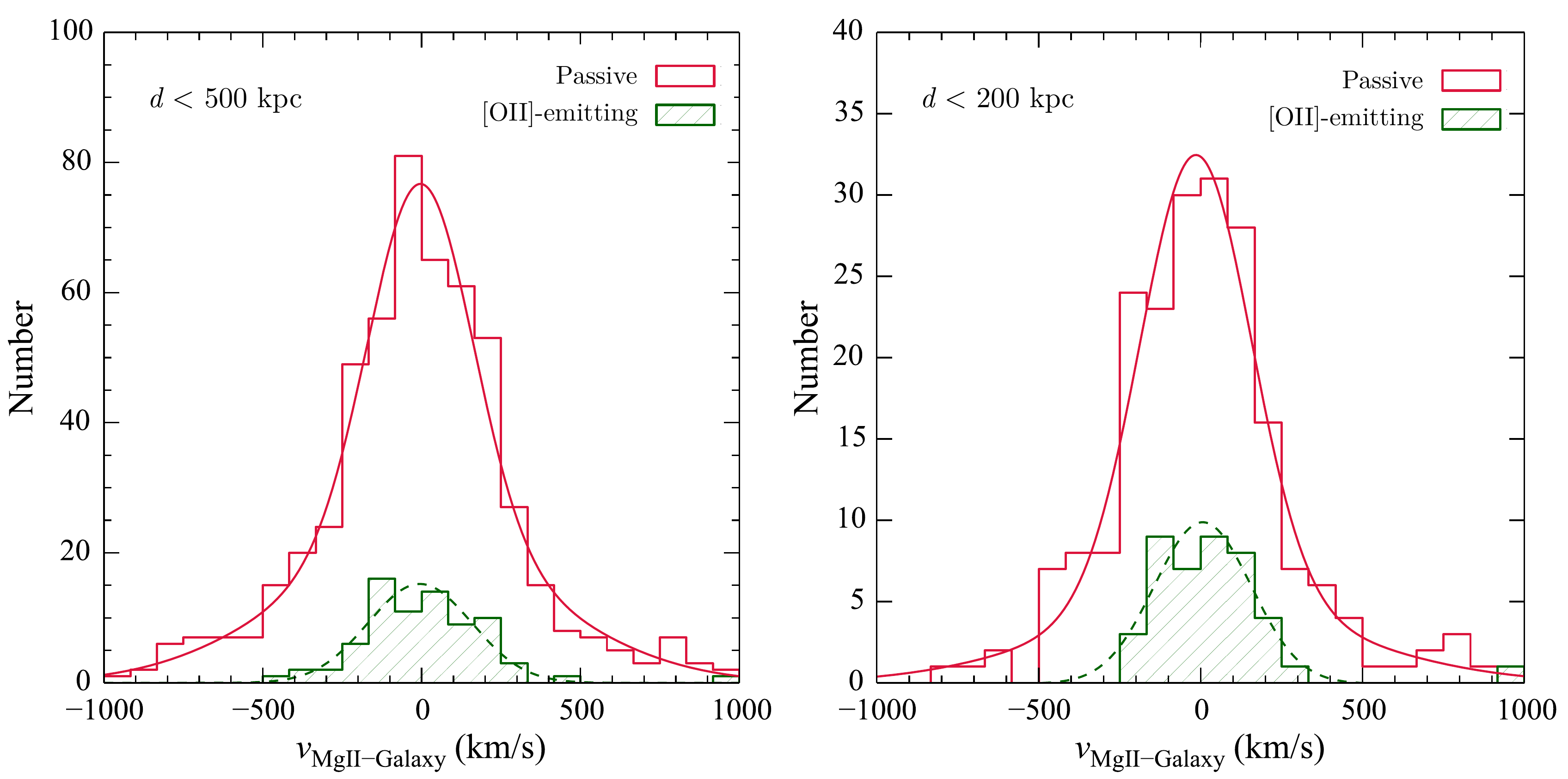}
\caption{Relative velocity distributions of \MgII absorbers with
  respect to the systemic redshifts of the LRGs.  Passive LRGs are
  shown in the red open histogram, while \OII-emitting LRGs are shown
  in the green hatched histogram.  The left panel includes all \MgII
  absorbers at $d<500$ kpc, and the right panel includes only those at
  $d<200$ kpc.  Mg\,II absorbing gas around \OII-emitting LRGs at
  $d<500$ (200) kpc is well characterized by a single Gaussian
  distribution centered at $\langle v_{\rm
    Mg\,\scriptsize{II}-Galaxy}\rangle = -5$ ($+6$) \kms\ with a
  dispersion of $\sigma_v = 167$ (150) \kms\ (green, dashed curve).
  Mg\,II absorbing gas around passive LRGs exhibits a similar
  distribution but with a substantial fraction, ($12\pm 1$)\% at
  $d<500$ kpc and ($6\pm 2$)\% at $d<200$ kpc, extended beyond
  $|v_{\rm Mg\,\scriptsize{II}-Galaxy}|=500$ \kms\ from the systemic
  redshifts of the galaxies.  Applying a double Gaussian profile to
  characterize the velocity distribution of Mg\,II gas around passive
  LRGs at $d<500$ (200) kpc leads to a narrow component centered at
  $\langle v_{\rm Mg\,\scriptsize{II}-Galaxy}\rangle = -3$ ($-15$)
  \kms\ and $\sigma_v = 163$ (170) \kms\ and a broad component
  centered at $\langle v_{\rm Mg\,\scriptsize{II}-Galaxy}\rangle =
  -17$ ($+6$) \kms\ and $\sigma_v = 415$ (453) \kms\ (red solid
  curve).  }
	\label{figure:vdiff}
\end{figure*}

To examine Mg\,II absorption properties in LRG halos, we first compare
the observed Mg\,II absorption strength with the projected distance
between the absorbing gas and the galaxy.  Figure \ref{figure:EWrho03}
shows the distribution of \ewr\ versus projected distance $d$ to the
LRGs.  We present the observations for passive and \OII-emitting LRGs
in separate panels for a direct comparison of the CGM properties
between galaxies with different star formation properties (see the
discussion in the previous section, \S\ 3.2).  We focus on the
subsample of LRGs with sufficiently high $S/N$ background QSO spectra
for detecting Mg\,II absorbers as weak as $W_0=0.3$ \AA.  A
significant fraction of the original SDSS LRG sample have
poorer-quality QSO spectra that result in upper limits exceeding 0.3
\AA.  As a result, these LRG--QSO pairs offer little/no constraints
for the underlying absorber strengths in LRG halos and are therefore
excluded from the panels for clarity.

Both panels in Figure \ref{figure:EWrho03} show that
%the number of LRG--QSO pairs increases with $d$ roughly as $d^2$
the LRGs with associated Mg\,II absorbers occupy a similar
\ewr\ versus $d$ space.  Although many LRGs show no detectable Mg\,II
absorbers of $\ewr\geq 0.3$ \AA, a non-negligible fraction of LRGs
show strong associated \MgII absorbers of $\ewr \sim 1$ \AA\ out to
$d=500$ kpc, the virial radii of LRG host dark matter halos.  While
strong \MgII absorbers are also found at large distances from QSOs
(e.g., Johnson \etal\ 2015), such flat $\ewr$ versus $d$ trend is in
stark contrast to known halo gas properties around $L_*$ galaxies.
These strong absorbers are only found around $L_*$ galaxies at $d\apl
60$ kpc, beyond which the observed Mg\,II absorbing strength rapidly
declines (e.g., Chen \& Tinker 2008; Chen \etal\ 2010a).  

%\begin{figure}
%\centering
%\includegraphics[scale=0.3]{mass_comp.pdf}
%\caption{Stellar mass distributions of \MgII-absorbing and
%  non-absorbing LRGs to illustrate the lack of mass dependence on the
%  incidence of \MgII absorbers in LRG halos. The {\it left} panel
%  displays the stellar mass distribution for passive LRGs, and the
%  {\it right} panel displays the stellar mass distribution for
%  \OII-emitting LRGs.  The red, open histograms are for LRGs without
%  an associated \MgII absorber at a 2-$\sigma$ upper limit of
%  $\ewr=0.3$ \AA, and the green, dashed histograms are for those with
%  associated \MgII absorbers.  For a baseline comparison, we include
%  the mass distribution for the full sample of 37621 LRGs from Figure
%  1 in orange, solid histograms.  We do not find a preference of \MgII
%  absorbers around low- or high-mass LRGs.}
%	\label{figure:masscomp}
%\end{figure}

As described in \S\ 2.1, the LRGs span a range in stellar mass from
$\log\,M_*/M_\odot<11$ to $\log\,M_*/M_\odot\approx 12$ (right panel
of Figure 1) with a mean of $\langle\,\log\,M_*/M_\odot\,\rangle=11.4$
and a dispersion of 0.2 dex.  It is possible that the relatively broad
range in stellar mass contributes to the observed flat trend in $\ewr$
versus $d$.  We perform two tests to examine whether this is a factor.
First, we compare the stellar mass distributions of \MgII-absorbing
and non-absorbing LRGs.
%Figure
%\ref{figure:masscomp} shows that for either passive (left panel) or
%\OII-emitting (right panels) LRGs, both \MgII-absorbing (green, dashed
%histograms) and non-absorbing (red, open histograms) LRGs share the
%same stellar mass distribution as the full sample (orange, solid
%histograms).  
We find that both \MgII absorbing and non-absorbing LRGs are well
characterized by a Gaussian distribution function with a mean of
$\langle\,\log\,M_*/M_\odot\,\rangle=11.4$ and a dispersion of 0.2
dex.  We therefore do not find a preference of \MgII absorbers around
low- or high-mass LRGs.  Next, we include the stellar mass scaling
relation found for $L_*$ galaxies by Chen \etal\ (2010b) and examine
whether the observed scatter in $\ewr$ versus $d$ is reduced.  The
result shows that including stellar mass scaling does not
improve/reduce the scatter in the observed $\ewr$ versus $d$ relation.
Both tests confirm that extended \MgII-absorbing gas does not depend
strongly on the mass of the LRGs.

%Several studies show that finding galaxy
%groups associated with strong \MgII absorbers may
%not be an uncommon occurrence.  The studies include
%\cite{Kacprzak:2010} of a \ewr = 1.8 $\rm \AA$ \MgII absorber,
%and \cite{Whiting:2006} of a \ewr\ $\approx 2.5\rm\AA$ \MgII absorber.
%In the group environment, the \ewr\ has a bimodal feature (either
%very high or low detection, ref:XX).  

At the same time, we also examine the velocity dispersion of the
detected absorbing gas around LRGs.  The left panel of Figure
\ref{figure:vdiff} shows the relative line-of-sight velocity
distributions of \MgII absorbers with respect to the systemic
redshifts of LRGs at $d<500$ kpc.  Following the presentation in
Figure \ref{figure:EWrho03}, we present the velocity distribution
separately for passive and \OII-emitting LRGs.  The velocity
distribution of \MgII absorbing gas around \OII-emitting LRGs can be
characterized by a single Gaussian distribution of mean velocity
difference $\langle v_{\rm Mg\,\scriptsize{II}-Galaxy}\rangle = -5$
\kms\ and dispersion $\sigma_v = 167$ \kms\ (green dashed curve).
However, the velocity distribution of \MgII absorbing gas around
passive LRGs without detectable \OII\ emission features appears to
have extended, high-velocity wings and is best represented by a double
Gaussian profile with a narrow component centered at $\langle v_{\rm
  Mg\,\scriptsize{II}-Galaxy}\rangle = -3$ \kms\ and $\sigma_v =163$
\kms\ and a broad component centered at $\langle v_{\rm
  Mg\,\scriptsize{II}-Galaxy}\rangle = -17$ \kms\ and $\sigma_v = 415$
\kms\ (red solid curve).  We find that 62 \MgII absorbers occur at
$|v_{\rm Mg\,\scriptsize{II}-Galaxy}| >500$ \kms\ from passive LRGs,
which constitute $(12\pm 1)$\% of the total \MgII-absorbing passive
LRG sample.

To evaluate whether the observed velocity dispersion vary with
projected distance, we consider only LRGs with a Mg\,II absorber found
at $d<200$ kpc from LRGs.  Of the 540 passive LRGs--Mg\,II absorber
pairs in Table 2, 208 are separated by $d<200$ kpc. Of the 76
\OII-emitting LRGs--Mg\,II absorber pairs, 42 are separated by $d<200$
kpc.  The right panel of Figure \ref{figure:vdiff} shows the
line-of-sight velocity distribution of \MgII absorbers relative to the
LRGs at $d<200$ kpc.  We find similar characteristics in the
velocity distribution of \MgII absorbing gas at smaller projected
distances from LRGs.  For \OII-emitting LRGs, a single Gaussian
function is sufficient to describe the line-of-sight gas motion with a
mean of $\langle v_{\rm Mg\,\scriptsize{II}-Galaxy}\rangle = +6$
\kms\ and dispersion $\sigma_v = 150$ \kms\ (green dashed curve in
Figure \ref{figure:vdiff}).  For passive LRGs, high-velocity \MgII
absorbers are also seen but at a reduced fraction.  We find that 13
out of 208 \MgII absorbers at $d<200$ kpc occur at $|v_{\rm
  Mg\,\scriptsize{II}-Galaxy}| >500$ \kms, which constitute
$(6\pm2)$\% of \OII-emitting LRGs with associated \MgII.  The best-fit
double Gaussian profile is characterized by a narrow component
centered at $\langle v_{\rm Mg\,\scriptsize{II}-Galaxy}\rangle = -15$
\kms\ and $\sigma_v =170$ \kms\ and a broad component centered at
$\langle v_{\rm Mg\,\scriptsize{II}-Galaxy}\rangle = +6$ \kms\ and
$\sigma_v = 453$ \kms\ (red solid curve in Figure \ref{figure:vdiff}).
The increasing fraction of high-velocity ( $|v_{\rm
  Mg\,\scriptsize{II}-Galaxy}| >500$ \kms) \MgII absorbers from
$d<200$ to larger distances may be understood by an increasing
fraction of contaminating random background or correlated absorbers
outside the LRG halos as the projected distance increases.

Figure \ref{figure:vdiff} shows that Mg\,II absorbing gas detected
around passive and \OII-emitting LRGs (particularly the absorbers
found at $d<200$ kpc) shares a similar line-of-sight velocity
dispersion of $\sigma_v\approx 165$ \kms.  The velocity field of
chemically-enriched gas in massive LRG halos does not exhibit
traceable dependence on the presence/absence of on-going star
formation in the galaxies.  Furthermore, we note that the mean halo
mass of LRGs is $\langle\,M_h({\rm LRG})\rangle\approx
10^{13.4}\,M_\odot$ (e.g.\ Mandelbaum \etal\ 2008; Gauthier \etal\
2009).  The expected line-of-sight velocity dispersion for virialized
gas in halos of $10^{13.4}\,M_\odot$ is $\sigma_h\approx 265$ \kms.
The observed velocity dispersion in Mg\,II absorbing gas is merely
60\% of what is expected from virial motion, namely $\sigma_v\approx
0.6\,\sigma_h$.  A similar result has also been seen by Zhu \etal\
(2014), who reported a velocity bias of $\sigma_{\rm \MgII}\approx
0.5\,\sigma_h$.  Such suppression in gas motion not only shows that
the gas is gravitationally bound to the LRG halo but also that
additional mechanisms are necessary to slow down the motion of these
absorbing clouds.

\subsection{Incidence and Covering Fraction of \MgII Absorbers in LRG halos}
\label{section:kappa}

\begin{figure}
\centering
\includegraphics[scale=0.46]{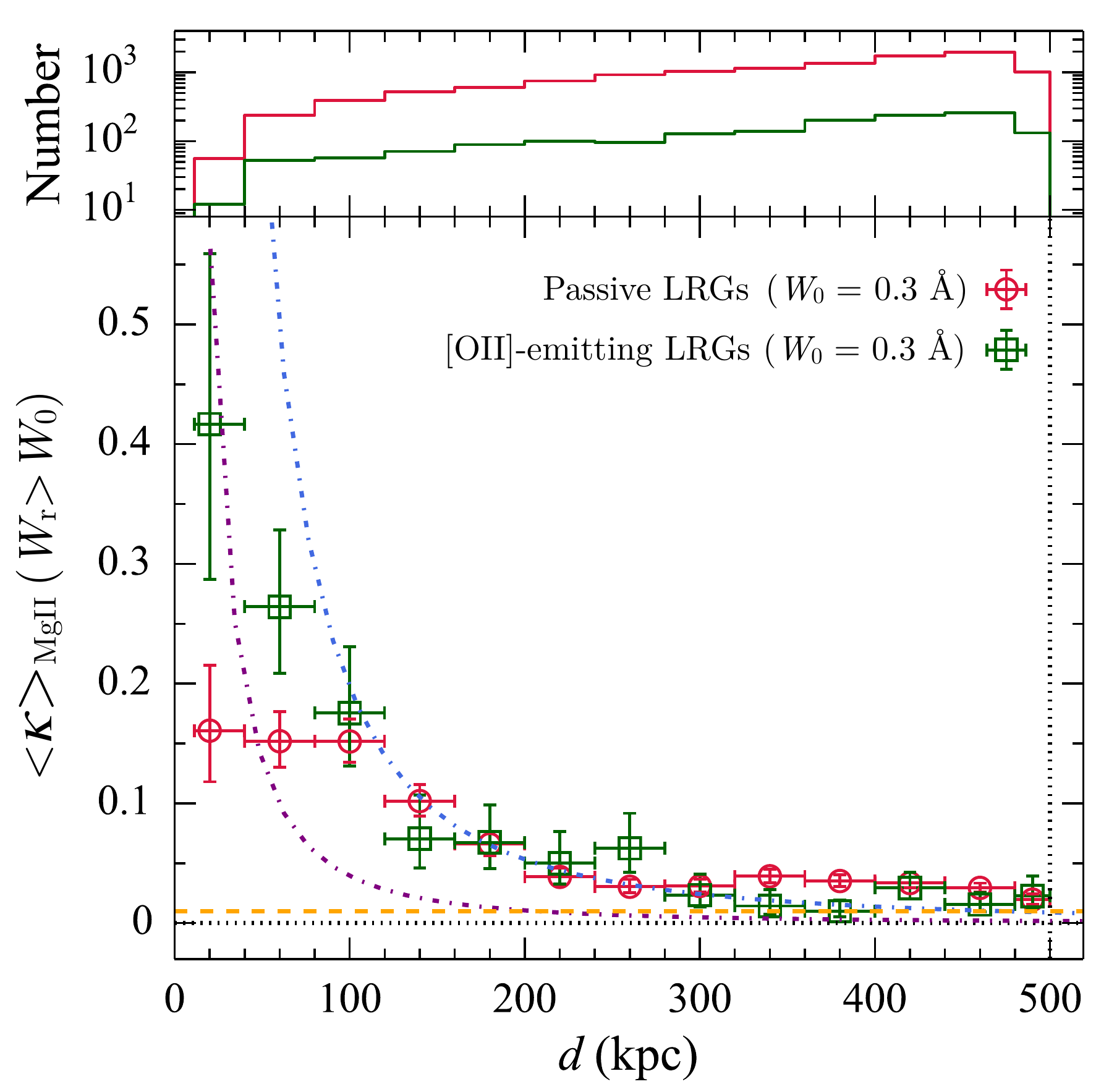}
\caption{Mean incidence (or covering fraction) of \MgII absorbing gas
  $\langle\kappa\rangle_\MgII$ versus projected distance $d$ for
  passive and \OII-emitting LRGs.  At the top, we show the number of
  LRGs that were considered in computing $\langle\kappa\rangle_\MgII$
  for each projected distance bin.  We adopt the LRG sample with
  sufficiently high $S/N$ background QSO spectra available for
  detecting Mg\,II absorbers as weak as $W_0=0.3$ \AA, and a bin size
  of 40 kpc.  The horizontal bars mark the full range of projected
  distance within each bin and vertical error bars represent the 68\%
  confidence intervals in the estimated gas covering fraction.  The
  vertical dotted line indicates the virial radius of a typical LRG
  halo.  We estimate the contribution due to random background
  absorbers that coincidentally occur within a velocity difference of
  $\Delta v = \pm1000\,{\rm km}\,{\rm s}^{-1}$ from the redshifts of
  the LRGs (orange dashed line; see Section \ref{section:kappa} for
  details).  The blue and purple dash-dotted curves indicate the
  expected maximum contributions to the observed incidence of
  \ewr\ $\geq$ 0.3 \AA\ absorbers from all and blue satellite
  galaxies, respectively (see Section \ref{section:satellites} for
  details).  }
	\label{figure:kappa03}
\end{figure}

A key quantity to characterize LRG halos is the covering fraction,
$\kappa_\MgII$, of chemically-enriched cool gas as revealed by the
presence of \MgII absorbers.  We employ a maximum likelihood analysis
to compute the best-fit $\kappa_\MgII$ and its associated
uncertainties as a function of projected distance $d$, following the
formalism described in \cite{Chen:2010a}.  The likelihood of observing
an ensemble of galaxies with $n$ showing associated \MgII and $m$
displaying upper limits is
\begin{equation}
  \begin{split}
  \mathcal{L}(\kappa_\MgII) = \langle\kappa\rangle_\MgII^n [1-\langle\kappa\rangle_\MgII]^m
  \end{split}
\end{equation}
We divide the LRGs into subsamples of different projected distance
intervals and compute best-fit $\kappa_\MgII$ and uncertainties for
each projected distance bin.  Figure \ref{figure:kappa03} shows the
estimated $\langle\kappa\rangle_\MgII$ versus $d$ in intervals of 40
kpc.  Error bars in $\langle\kappa\rangle_\MgII$ represent the 68
per cent confidence interval based on the likelihood function.  The
number of LRGs in each projected distance interval is shown at the top
of Figure \ref{figure:kappa03}.  For accurate estimates of gas
covering fraction, we consider only those LRGs with sufficiently high
$S/N$ background QSO spectra available for detecting Mg\,II absorbers
as weak as $W_0=0.3$ \AA.  For investigating possible dependence of
the incidence of chemically-enriched halo gas on star formation
activities, we consider passive and \OII-emitting LRGs separately.

We also estimate contamination due to background structures along the
line of sight.  Following \cite{Gauthier:2010}, we estimate the
incidence of random background \MgII absorbers within a redshift
interval of $\Delta\,z=0.01$, corresponding to a velocity interval of
$\Delta\,v=\pm 1000$ \kms, based on the mean number density of \MgII
absorbers with $\ewr\ge 0.3$ \AA\ from Nestor \etal\ (2005).  The
$\approx 1$ per cent contribution as shown in Figure
\ref{figure:kappa03} (dashed curve) suggests that random background
absorbers have negligible impact on the observed covering fraction of
\MgII absorbing gas at $d\apl 200$ kpc, but contribute a significant
portion of \MgII absorbers found at larger distances.

Figure \ref{figure:kappa03} displays a number of interesting features.
First, the covering fraction of \MgII absorbers is significantly
elevated in the inner halos with a mean covering fraction ranging from
$\langle\kappa\rangle_\MgII\approx 15$\% at $d\apl 120$ kpc for
passive LRGs to as high as $\langle\kappa\rangle_\MgII\approx 40$\% at
$d\apl 40$ kpc for \OII-emitting LRGs.  This is in contrast to an
overall covering fraction of $\langle\kappa\rangle_\MgII\approx 5$\%
at $d<500$ kpc.  In addition, $\langle\kappa\rangle_\MgII$ remains
flat at $\langle\kappa\rangle_\MgII=15$\% at $d\apl 120$ kpc from
passive LRGs, while it increases steadily with decreasing $d$ from
\OII-emitting LRGs.  Beyond $d\approx 120$ kpc, both passive and
\OII-emitting LRGs show consistent $\langle\kappa\rangle_\MgII$ to
within measurement uncertainties, which gradually declines and merges
into the background at $d>300$ kpc.
%Within $d<80$
%kpc, we identify \MgII absorption systems with \ewr $\geq 0.3\,\rm
%\AA$ around 19 of 65 LRGs with \OII\ and 45 of 293 LRGs without \OII,
%indicating a mean covering fraction of $\langle \kappa \rangle \approx
%0.29^{+0.06}_{-0.05}$ and $0.15^{+0.02}_{-0.02}$, respectively.  That
%is, the LRGs with \OII\ sample yields a $\sim 2$ times larger covering
%fraction than LRGs without \OII\ sample at $d<80$ kpc.  Interestingly,
%while the incidence of LRGs with \OII\ declines steeply with
%increasing $d$ at $d<80$ kpc, the covering fraction is constant for
%passive LRG sample.  At projected distances beyond $\apg$ 80 kpc, the
%covering fraction shows no significant difference between the two
%samples.  The mean covering fraction of two samples declines from
%$\langle \kappa \rangle \approx 0.16$ at $d=$ 80 kpc to $\langle
%\kappa \rangle \approx 0.05$ at $d=$ 200 kpc and remains at the same
%level of $\langle \kappa \rangle \approx 0.03$ out to $d=$ 500 kpc.

Recall the finding of \cite{Chen:2008} that typical $L_*$ galaxies
possess extended \MgII absorbing gas with an empirical gaseous radius
of $R_{\rm gas} = 130\,(L_B/L_{B_*})^{0.35\pm0.05}\, {\rm kpc} $.  For
a mean rest-frame $B$-band magnitude of $M_{\rm B}=-22.0$ of the LRGs,
corresponding to $3.6\,L_*$ for $M_{\rm B*}=-20.6$ from Cool
\etal\ (2012), we infer $R_{\rm gas} \approx 206 \,{\rm kpc}$ and find
a mean covering fraction of $\langle \kappa \rangle_\MgII =
14^{+2.0}_{-0.4}$\% at $d<R_{\rm gas}$ from \OII-emitting LRGs and
$\langle \kappa \rangle_\MgII = 11^{+1.0}_{-0.2}$\% at $d<R_{\rm gas}$
from passive LRGs.  The covering fraction is significantly smaller
than what is measured for both red and blue $L_*$/sub-$L_*$ galaxies
at $\langle \kappa \rangle_\MgII = (70\pm 10)$\% \citep{Chen:2010a},
confirming that the incidence of cool gas declines steeply with
increasing halo mass for halos of $M_h\apg\,10^{12}\,M_\odot$.

While such declining trend is expected in theoretical models that
attribute the observed \MgII absorbers to infalling gas from either
thermally unstable hot halos or the intergalactic medium (IGM)
\citep[e.g.,][]{Maller:2004, Keres:2009}, the large clustering
amplitudes found for \MgII absorbers (e.g., Bouch\'e \etal\ 2006;
Lundgren \etal\ 2009; Gauthier \etal\ 2009) also indicate that the
incidence and covering fraction of \MgII absorbing gas is non-zero in
high-mass halos (e.g., Tinker \& Chen 2008).  Indeed, our study based
on an unprecedentedly large LRG--QSO pairs has led to a definitive
detection of chemically-enriched cool gas around quiescent galaxies at
a level that conclusively rules out zero covering fraction.

To better understand the origin of chemically-enriched cool gas in
predominantly quiescent halos, we examine the spatial distribution of
\MgII absorbing gas relative to the orientation of the host LRGs.
This is motivated by the expectations that star formation driven
outflows are likely to proceed along the minor (spin) axis of a disk
galaxy and that accretion is likely to proceed along the major axis.
The former has been seen by a number of previous studies (e.g.,
Bordoloi \etal\ 2011) but is an unlikely scenario for explaining the
\MgII absorbers found around LRGs due to the observed low SFR.  At the
same time, it has been shown both in simulations and in observations
that elliptical galaxies preferentially have their major axes aligned
with the filamentary structures (e.g., Arag\'on-Calvo \etal\ 2007;
Tempel \etal\ 2013), where accretion of intergalactic gas and
satellites originates (e.g., Faucher-Gigu\`ere \& Kere\v{s} 2011;
Fumagalli \etal\ 2011; Tempel \etal\ 2015).

\begin{figure*}
\centering
\includegraphics[scale=0.55]{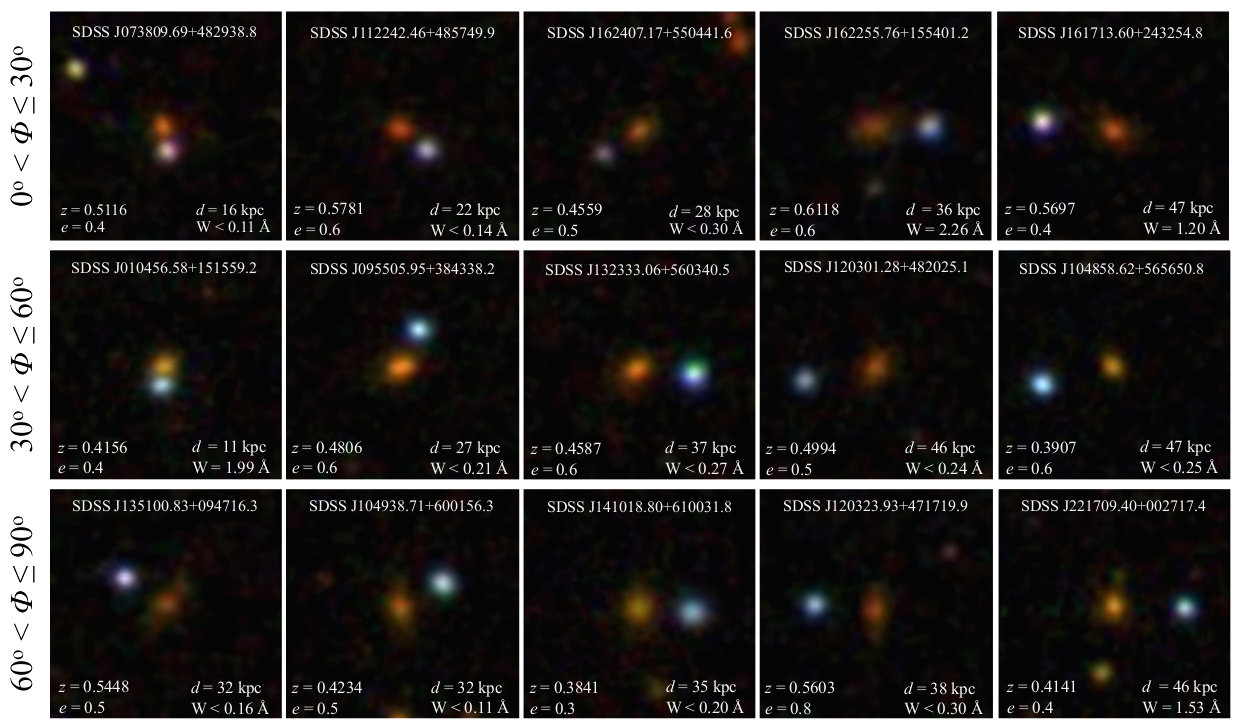}
\caption{Images of 15 LRG-QSO pairs with $d<50$ kpc to demonstrate
  visually that the accuracy of ellipticity ($e$) and azimuthal angle
  ($\Phi$) measurements from SDSS is sufficient for the subsequent
  $\Phi$-dependence study.  Each panel is 150 kpc on a side.  The LRG
  is placed at the center and the QSO appears as a blue compact source
  near the LRG.  The projected distance of each LRG is shown in the
  lower-right corner together with the observed constraint for
  associated \MgII absorbers (measurement or 2-$\sigma$ upper limits).
  The redshift and ellipticity of the LRG are included in the
  lower-left corner.  The top, middle, and bottom rows display
  examples of LRGs with consistent measurements of $\Phi$ from their
  $r$- and $i$-band images in the range of $0^\circ < \Phi \leq
  30^\circ$, $30^\circ < \Phi \leq 60^\circ$, and $60^\circ < \Phi
  \leq 90^\circ$, respectively.}
	\label{figure:image}
\end{figure*}

\subsection{Angular Distribution of \MgII Absorption Relative to Galaxy Major Axis}

To examine the azimuthal dependence of $\langle\kappa\rangle_\MgII$,
we make use of the position angle (P.A.) and ellipticity ($e$)
measurements, and associated measurement uncertainties, of each galaxy
from the SDSS database.  We refer the readers to \cite{Stoughton:2002}
for details on the SDSS image processing algorithm.  Briefly, the
position angle and ellipticity of each LRG were determined based on a
two-dimensional surface brightness profile fit of a deVaucouleurs
model to individual SDSS images.  Uncertainties in the best-fit
parameters depend on a combination of factors including the depth of
the images and the size of the galaxy relative to the size of the
point spread function of the image.  For LRGs at $z=0.4-0.7$, SDSS
$r$- and $i$-bands serve as the most sensitive bandpasses for
recording their surface brightness profiles.  We define the azimuthal
angle $\Phi$ of each QSO-LRG pair as the angle that extends from the
observed major axis of the LRG to the location of the background QSO
sightline.  Following this definition, a QSO sightline that occurs
along the major axis of the galaxy has $\Phi=0^\circ$ and a QSO that
occurs along the minor axis has $\Phi=90^\circ$.

\begin{figure*}
\includegraphics[angle=0,scale=0.45]{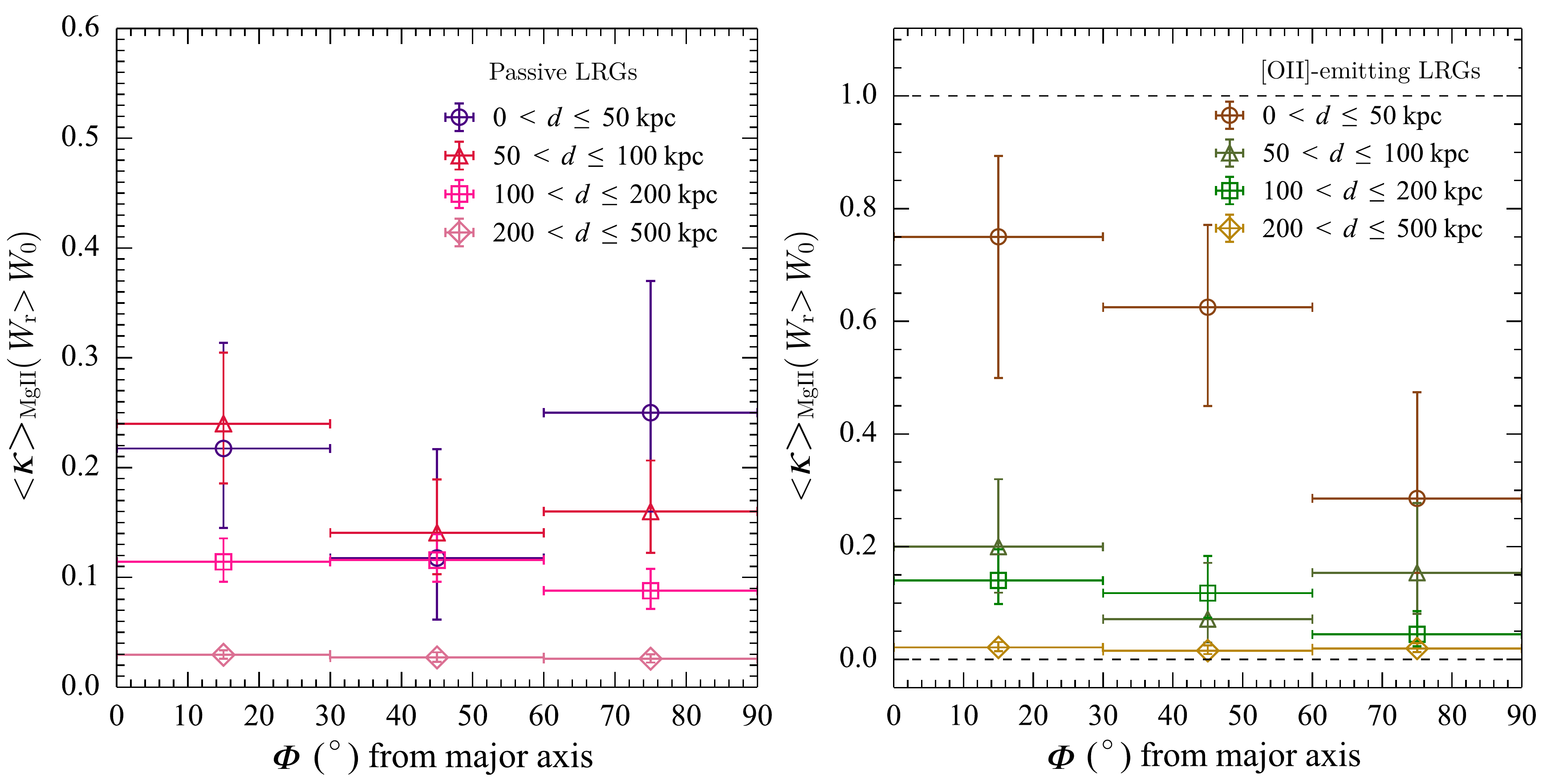} 
\caption{Dependence of $\langle\kappa\rangle_\MgII$ on the azimuthal
  angle in different projected distance intervals for passive ({\it
    left}) and \OII-emitting LRGs ({\it right}).  The azimuthal angle
  is measured with respect to the major axis of the LRGs.  Circles
  represent LRGs within projected distance $d$ of $\rm 0 < {\it d}
  \leq 50\ kpc$ from a background QSO sightline, triangles represent
  $\rm 50 < {\it d} \leq 100\ kpc$, squares represent $\rm 100 < {\it
    d} \leq 200\ kpc$ and diamonds represent $\rm 200 < {\it d} \leq
  500\ kpc$.  The horizontal error bars show the full range in
  projected distance of each bin and vertical error bars represent the
  68\% confidence interval.  Note that we have zoomed in along the
  y-axis in the left panel in order to better illustrate the range of
  observed $\langle\kappa\rangle_\MgII$ for the passive galaxies.}
	\label{figure:PA}
\end{figure*}

To ensure high confidence in the results of of our azimuthal
dependence investigation, we first consider only LRGs that show
consistent measurements in both P.A.\ and $e$ from SDSS $r$- and
$i$-band imaging data.  In addition, we restrict our sample for this
study to those LRGs with a measured ellipticity consistently greater
than $e=0.2$ in both $r$- and $i$-bands.  This criterion removes 2545
LRGs from the total sample of 13330 LRGs.  We then divide the
remaining QSO-LRG pairs into different bins in $\Phi$.  The bin size
is chosen to be larger than the uncertainties in $\Phi$, which are
evaluated based on the consistency of P.A.\ measurements from $r$- and
$i$-band images.  We find a typical error of $<10^\circ$ for the P.A,
and selected a bin size of $\Delta\,\Phi=30^\circ$.  In Figure
\ref{figure:image}, we show 15 (out of 70 total) LRG-QSO pairs with
$d< 50$ kpc to illustrate typical cases with azimuthal angle falling
in three bins: $0^\circ<\Phi\leq30^\circ$ (top panels),
$30^\circ<\Phi\leq60^\circ$ (middle panels), and
$60^\circ<\Phi\leq90^\circ$ (bottom panels).  A visual inspection of
Figure \ref{figure:image} confirms that measurements of $\Phi$ are
sufficiently accurate for the adopted bin size in $\Phi$.

In Figure \ref{figure:PA}, we show $\langle\kappa\rangle_\MgII$ as a
function of $\Phi$ in different projected distance bins.  Passive LRGs
are shown in the left panel and \OII-emitting LRGs are shown in the
right panel.  At $d\geq$50 kpc, we find no strong dependence of
$\langle\kappa\rangle_\MgII$ on $\Phi$ for either passive or
\OII-emitting LRGs.  While there is little azimuthal angle preference
for passive LRGs at $d\leq$ 50 kpc, we find a modest ($\approx 50$\%)
enhancement of \MgII absorption closer to the major axis ($\Phi\apl
60^\circ$) of \OII-emitting LRGs.  The excess of \MgII covering
fraction decreases with increasing $\Phi$ and becomes consistent with
that of passive LRGs at $\Phi>60^\circ$.  Recall that we have shown in
Figure \ref{figure:kappa03} the overall enhancement of covering
fraction for \OII-emitting LRGs as compared to passive LRGs at $d\leq$
80 kpc.  We find this difference in $\langle\kappa\rangle_\MgII$ is
likely driven by an elevated incidence of \MgII absorbing gas at small
azimuthal angles.  The difference in the observed azimuthal dependence
between \OII-emitting and passive LRGs suggests that additional
sources are responsible for the observed \MgII absorbing gas around
[OII]-emitting galaxies.  We will discuss this further in \S\ 4.4
below.

\section{Discussion}
\label{section:discussion}

Our study based on an unprecedentedly large sample of LRG-QSO pairs
has led to a definitive detection of chemically-enriched cool gas
around massive, quiescent galaxies at a level that conclusively rules
out zero covering fraction of cool gas in these massive halos.  The
result is based on a survey of 37621 LRG-QSO pairs with projected
separations of $<500$ kpc, which yielded a sample of 13330 LRGs at
$z\approx 0.4-0.7$ with sensitive background QSO spectra available for
constraining the presence/absence of \MgII absorbers of $\ewr\ge 0.3$
\AA.  Roughly 13\% of these LRGs exhibit \OII\ emission features that
indicate a mean on-going star formation rate of ${\rm SFR}\sim
0.8\,M_\odot\,{\rm yr}^{-1}$ among otherwise old (${\rm age}\apg 1$
Gyr) stellar populations.  The remaining 87\% of the LRGs exhibit no
trace of on-going star formation with 2-$\sigma$ upper limit of ${\rm
  SFR}\apl 0.1\,M_\odot\,{\rm yr}^{-1}$ and a mean stellar age
$\apg 1$ Gyr.  Both passive and \OII-emitting LRGs share a very
similar distribution in stellar mass with a mean of
$\langle\,\log\,(M_*/M_\odot)\,\rangle \approx 11.4$ and a dispersion
of 0.2 dex.

Strong \MgII absorbers are found at $d<500$ kpc from both passive and
\OII-emitting LRGs with a mean gas covering fraction of
$\langle\kappa\rangle_\MgII\approx 5$\%.  While strong \MgII absorbers
continue to be found at distances as large as the virial radius in the
halos, the mean gas covering fraction declines rapidly with increasing
$d$.  The mean covering fraction of \MgII absorbers increases to
$\langle\kappa\rangle_\MgII\approx 11-14$\% within the fiducial
gaseous radius $R_{\rm gas}\approx 200$ kpc inferred for super-$L_*$
galaxies from Chen \& Tinker (2008), and continues to increase to
$\langle\kappa\rangle_\MgII\apg 15$\% at smaller radii at $d\apl 120$
kpc.  At $d<80$ kpc, the observed gas covering fraction around
\OII-emitting LRGs is twice of what is seen around passive LRGs.  No
clear dependence on stellar mass is found for the observed \MgII
absorption properties.  These results confirm and significantly
improve upon earlier reports for the presence of cool halo gas around
LRGs by Gauthier \etal\ (2009, 2010).

In addition to constraining the incidence and covering fraction of
\MgII absorbers in LRG halos, we have also further examined the
kinematic and spatial distribution of \MgII absorbing gas relative to
the galaxies.  An intriguing finding is that the observed velocity
dispersion of Mg\,II absorbing gas relative to either passive or
\OII-emitting LRGs is merely 60\% of what is expected from virial
motion in these massive halos (\S\ 3.2 \& Figure 4), which is similar
to what has been previously reported by Zhu \etal\ (2014).
Furthermore, we have also investigated possible azimuthal dependence
in the incidence and covering fraction of \MgII absorbers of $\ewr\ge
0.3$ \AA.  While no apparent trend is seen for passive LRGs at all
radii, a surprising result is a modest enhacement in the gas covering
fraction along the major axis of \OII-emitting LRGS at $d<50$ kpc
(\S\ 3.3 \& Figure \ref{figure:PA}).  This is opposite of what was
found for star-forming galaxies at $z\approx 0.7$ by Bordoloi
\etal\ (2011).

The observed suppression in the velocity dispersion of \MgII absorbing
gas around both passive and \OII-emitting LRGs, together with an
elevated \MgII gas covering fraction along the major axis of
\OII-emitting LRGs at $d<50$ kpc, provides important insights into the
origin of the observed chemically-enriched cool gas in LRG halos.
Here we discuss whether/how different scenarios are compatible with
these findings.

\subsection{Hot Winds due to AGN or Evolved Stars}

We first consider the scenario of hot winds driven by either evolved
stars or AGN as a primary driver to pollute the LRG halos with heavy
elements.  While the observed low SFR and a predominantly old stellar
population rules out a strong influence of young starburst driven
winds on the halo gas of LRGs, the influence of AGN feedback can be
dominant (e.g., McNamara \& Nulsen 2007, 2012).  In particular,
radio-mode feedback has been invoked in galaxy formation models to
suppress star formation in massive halos (e.g., Croton \etal\ 2006).
In addition, a cross-comparison between SDSS LRGs and FIRST radio
sources has identified $\approx 3$\% of the LRGs hosting radio-loud
AGN (e.g., Sadler \etal\ 2007), and efforts in search of fainter radio
emission in the remaining LRGs based on median stacks of FIRST images
have continued to uncover radio signals at a level of a few
$\times\,10\,\mu$Jy (e.g., Hodge \etal\ 2008, 2009).  Continuing
detections of radio fluxes in LRGs as the sensitivities of the
searches improve suggests that nearly all LRGs harbor an active
nucleus but with varying radio power.

The presence of AGN in these LRGs also provides a natural explanation
for the observed high [N\,II]\,/\,\Halpha\ emission ratio in the
stacked LRG spectra presented in Figure \ref{figure:spec_med} (e.g.,
Johnston \etal\ 2008; Hodge \etal\ 2008).  It is therefore
reasonable to expect that the LRG halos are being regulated by AGN
winds that provide an additional heating source.

In contrast, the observed low [O\,III]\,/\,[O\,II] ratios in
\OII-emitting LRGs resemble the spectra of LINER-like galaxies, rather
than star-forming regions or Seyfert galaxies (e.g., Yan \etal\ 2006).
The observed LINER-like spectra can be explained by photo-ionization
due to post-asymptotic giant branch (post-AGB) stars (e.g., Binette
\etal\ 1994).  This is motivated by recent observations that have
uncovered spatially extended LINER signals in passive red galaxies
with a surface brightness profile shallower than $r^{-2}$ (e.g., Sarzi
\etal\ 2006, 2010; Yan \& Blanton 2012; Singh \etal\ 2013).  The
shallow, extended surface brightness profiles are inconsistent with
the gas being ionized by a central point source, but indicate a
spatially distributed ionizing source.

However, connecting the observed cool gas revealed by \MgII absorption
with AGN/stellar winds remains challenging, particularly because of
the suppressed velocity dispersion of \MgII absorbing gas relative to
the LRGs.  Recall from Figure \ref{figure:vdiff} and \S\ 3.2 that the
observed velocity dispersion of Mg\,II absorbing gas relative to
either passive or \OII-emitting LRGs is merely 60\% of what is
expected from virial motion in these massive halos.  One would expect
that including AGN/stellar winds would further stir up halo gas motion
(e.g., Johnson \etal\ 2015), increasing/maintaining the velocity
dispersion rather than suppressing it.  We therefore find this
scenario to be an unlikely explanation of the observed \MgII absorbers
in LRG halos.

%TODO: other outflows.
%\cite{Liu:2013} finds observational evidence that AGN
%can also drive outflows.  Measure the OIII luminosity of LRGs, 
%which can infer the luminosity of AGN.  Compare to the
%paper and say the luminosity doesn't match. \\
%Ancient outflow: DavÃ©, Romeel has been working on this. 
%We can constrain from the expected life time of MgII cloud,
%driven outflow velocity, and probably stellar population 
%synthesis model (old stellar population with a SF on top; e.g.,
%jar's paper showing LRGs are very old). \\ 

\subsection{Environmental Effects}%: Contributions from Satellite Galaxies}
\label{section:satellites}

Next, we consider possible environmental effects that may contribute
to the observed \MgII absorbers in LRGs halos.  The large mean bias
found for LRGs (e.g., Padmanabhan \etal\ 2007; Gauthier \etal\ 2009)
indicate not only that these galaxies reside in massive halos but also
that they reside in relatively more overdense environment.  First, we
consider gas-rich satellites that could contribute to some fraction of
the observed \MgII absorbers, if the satellites can retain a
significant fraction of their gas.

Under the assumption that the gas content of satellite galaxies
remains intact in LRG halos, we estimate the expected {\it maximal}
contribution to the covering fraction of \MgII absorbing gas from
these satellites.  Following \cite{Gauthier:2010}, we first adopt the
subhalo mass function from \cite{Tinker:2010b}.  The subhalo mass is
defined as the mass at the time of accretion and therefore is suitable
for calculating the intact gaseous halo.  Next, we adopt the gaseous
radius at $\ewr \geq 0.3$ \AA, $R_{\rm gas}$, and a mean covering
fraction of $\kappa_{\rm gas}$ within $R_{\rm gas}$ for a given
subhalo mass ($M_{\rm sub}$) from \cite{Tinker:2008}.  Then we adopt
the surface mass density profile of satellite galaxies from
\cite{Budzynski:2012}, which is characterized by a projected
Navarro$-$Frenk$-$White profile with a concentration parameter
$\langle c \rangle \equiv \langle r_{\rm vir}/ r_{\rm s} \rangle
\approx 2.6$ from , where $r_{\rm vir}$ is the viral radius and
$r_{\rm s}$ is the scale radius.  The best-fit concentration parameter
was found to be nearly independent of mass and a factor of two lower
than what is found for the dark matter halo.  Given a host halo mass
$M_{\rm host}$ (in this case, $10^{13.4} M_\odot$), the covering
fraction of subhalos as a function of projected distance $d$ is then
computed according to
\begin{equation}
\begin{aligned}
\kappa_{\rm sub}(M_{\rm host}, d) = &\frac{f(d,M_{\rm host})}{\pi R^2_{\rm vir}(M_{\rm host})}
 \int d M_{\rm sub} \,n(M_{\rm sub}| M_{\rm host}) \\
& \times \pi R^2_{\rm g}(M_{\rm sub}, W_{\rm r}=0.3)
\kappa_{\rm g}(M_{\rm sub}) 
\end{aligned}
\end{equation}
where $f(d,M_{\rm host})$ is the probability of having subhalos at $d$
%that takes into account the subhalo distribution 
and $n(M_{\rm sub}| M_{\rm host})$ is the subhalo mass function from
\cite{Tinker:2010b}.  The estimated, {\it maximal} covering fraction
of \MgII absorbers versus projected distance is shown in Figure
\ref{figure:kappa03} (blue dashed-dotted curve).  Our calculation
demonstrates that if satellite galaxies can retain their gas, then
they can fully account for the observed \MgII covering fraction.

Many studies have shown that galaxies in denser environments tend to
have a higher fraction of red galaxies at $z\apl1$
\citep[e.g.,][]{Gerke:2007,Skibba:2009,Smith:2012,Kovac:2014},
indicating that the star formation has been shut down either due to
gas exhaustion or removal by environmental effects.  Quantitatively
speaking, the red satellite fraction decreases from $\sim 80\, \%$ at
projected radius $\apl 100 \rm\, \rm kpc$ from $L_*$ satellite
galaxies to $\sim 70\,\%$ at about the virial radius
\cite[e.g.,][]{Hansen:2009,Prescott:2011}.  Here we consider only
$L_*$ halos because they are expected to be the dominant contributor
to the \MgII covering fraction from a halo occupation analysis
\citep[][]{Tinker:2008,Tinker:2010a}.  If we further restrict the blue
(and therefore gas rich) satellite fraction to be 20\,\% and assume
that these blue satellites can retain their gaseous halos, then the
expected, maximal blue satellite contribution is shown as the purple
dashed-dotted curve in Figure \ref{figure:kappa03}.  We find that blue
satellites alone cannot account for the observed 15\% covering
fraction of \MgII absorbing gas at $d\approx 100$ kpc from the LRGs
but could be a main contributor to the incidence of gas at $d\apl 40$
kpc.

Incidentally, the spatial distribution of satellites is found to be
aligned with the major axis of the brightest galaxies in groups
\citep[e.g.,][]{Yang:2006,Donoso:2006,Wang:2008}.  While the effect is
more subtle for blue satellites than red satellites, \cite{Yang:2006}
found that blue satellites along the major axis are $\sim$ 25 \% more
abundant than along the minor axis.  This is qualitatively consistent
with the trend found in the azimuthal dependence of
$\langle\kappa\rangle_\MgII$ in Figure \ref{figure:PA}.

However, ram pressure and tidal stripping are expected to be effective
in removing gas from satellite galaxies
\citep[e.g.,][]{Gunn:1972,Balogh:2000, Kawata:2008}.  These dynamical
processes should re-distribute chemically-enriched cool ISM and halo
gas of blue satellites to larger distances, consequently further
suppressing the incidence of cool gas due to satellites in inner LRG
halos.  This expectation is qualitatively consistent with the observed
flat distribution of $\ewr$ versus projected distance.  The presence
of strong Mg\,II absorbers out to the virial radius also suggests that
these LRGs possibily reside in a group/cluster environment
\citep[e.g.,][]{Whiting:2006,Kacprzak:2010,Gauthier:2013}.  A
remaining caveat is to explain the suppressed velocity dispersion of
\MgII gas around LRGs.  In summary, if red satellites have their gas
removed, then satellites do not explain the \MgII covering fraction,
unless the cool gas survives in the halo after removal.

\subsection{Condensing Cool Clouds due to Thermal Instabilities}

Next, we consider a third possibility of the observed \MgII absorbers
arising in cool clouds that are condensing out of thermally-unstable
hot halos around the LRGs.  A two-phase medium was first considered by
Mo \& Miralda-Escud\'e (1996) for explaining the observed QSO
absorption systems as cool clouds in pressure equilibrium with the hot
halo, which was subsequently expanded by Maller \& Bullock (2004) to
include multi-phase cooling for understanding the formation and
survival of high-velocity clouds found in the Milky Way halo.  These
earlier analytic models have relied on a simple hypothesis that
thermal instabilities would develop if the cooling time ($t_{\rm
  cool}$) is comparable to or smaller than the dynamical time ($t_{\rm
  ff}$) of the gas.  However, recent numerical simulations have
provided more detailed insights into the process of forming a
multi-phase medium (e.g., McCourt \etal\ 2012; Sharma \etal\ 2012).
It has been shown that, in fact, a multi-phase medium starts to develop
when the cooling time is $3-10$ times the free-fall time (Sharma
\etal\ 2012).  Indeed, multi-phase gas has been observed in clusters
and nearby ellipticals with extended nebular emission from relatively
cool gas embedded in hot, x-ray emitting halos (e.g., Werner
\etal\ 2014), and the gas in those cooling clusters and ellipticals is
found to satistfy the criterion of $t_{\rm cool}/t_{\rm ff}\apl 10$
(Voit \etal\ 2015a,b).

Direct observations to distinguish between the presence and absence of
a multi-phase medium around the LRGs in our sample are beyond the
reach of current-generation facilities.  Nevertheless, we expect that
these LRGs to be surrounded by a hot halo, given that the
Sunyaev-Zel'dovich decrement in the cosmic microwave background
radiation has been detected in stacks of 148 GHz maps of higher-mass
LRGs with $M_h\sim 10^{14}\,M_\odot$ (Hand \etal\ 2011).  Under the
multi-phase cooling hypothesis, we expect to see cool clouds form
within a cooling radius $R_c$ where thermal instabilities occur.
Observations of $z\approx 0.5$ galaxy clusters indicate that $R_c$
occurs between $1/3$ and $2/3$ of the virial radius (e.g., Voit
\etal\ 2015a).  For LRGs in our sample, this corresponds to a cooling
radius of $160-320$ kpc.  Condensing cool clouds may explain the rapid
decline in $\langle\kappa\rangle_\MgII$ at $d>120$ kpc, although they
cannot explain the strong \MgII absorbers detected in the outer halos
near the virial radius.

A natural expectation for condensed cool clouds traveling through a
hot halo is a ram-pressure drag force that would slow down the cloud
motion.  If the clouds are not sufficiently massive, then we expect to
observe significant deceleration.  This provides a physical model for
explaining the observed suppression of velocity dispersion between
\MgII absorbers and LRGs in Figure \ref{figure:vdiff} and places a
maximum limit on the cloud mass.  Following Maller \& Bullock (2004),
we compute the mass limit using their Equation (40),
\begin{equation}
m_{\rm cl}\approx 5.1\times 10^4\,M_\odot\,T_6^{-3/8}(\Lambda_z\,t_8)^{1/2},
\end{equation}
where $T_6$ is the halo gas temperature in units of $10^6\, \rm K$,
$\Lambda_z$ is the cooling parameter that varies with the gas
metallicity $\rm Z_g$ and $t_8 = t_f(c_h)/8\, {\rm Gyr}$ is the halo
formation time that depends on the halo concentration $c_h$.  We
estimate $T\sim 6\times10^6\,\rm K$ assuming isothermal hot gas for
the LRG halos, and $t_8 \sim 8.9\,\rm Gyr$ using $c_h\sim10$ from the
halo mass--concentration relation \citep[e.g.,][]{Mandelbaum:2008}.
We find $m_{\rm cl}=2.8\times 10^4\,M_\odot$ for a gas metallicity of
$Z_g = 0.1 \, Z_\odot$, and $m_{\rm cl}=4.9\times 10^4\,M_\odot$ for
$Z_g = Z_\odot$.  Note that for clouds of this low mass, it would
require $n_{\rm cl}\sim 20$ of these to make up the total observed
absorption strength per sightline (see Chen \etal\ 2010a).  This is
consistent with the previous finding that strong \MgII absorbers
identified in moderate-resolution spectra are routinely resolved into
multiple components in high-resolution spectra, with $\ewr$ roughly
proportional to the number of components in the system (e.g.,
Petitjean \& Bergeron 1990 ; Prochter \etal\ 2006 ).  However, the
inferred maximum cloud mass for the ram-pressure drag to be dominant
also indicates that the clouds would be likely to evaporate due to
thermal conduction of the hot halo in $\tau_{\rm evap}\sim 100$ Myr
from Equation (35) of Maller \& Bullock (2004).  Taking into account a
prolonged infall time of $\tau_{\rm infall}\approx 400-600$ Myr for
clouds formed at $R_c$ from Equation (43) of Maller \& Bullock (2004)
due to ram-pressure drag, we conclude that only clouds formed at
$d<100$ kpc will be able to reach the center.

\subsection{Accretion along Filaments}

Recent simulations have shown that galaxies of all mass acquire most
of their baryonic mass through filamentary accretion from the IGM
\citep[e.g.,][]{Keres:2009, Stewart:2011}.  In these narrow, dense
streams, the cool gas would never be shock heated to the virial
temperature and can penetrate deep toward the central regions of dark
matter halos.  Given that LRGs reside in overdense regions, a sizable
fraction of central LRGs may be at the focus of cold accretion through
filaments.  In the cold accretion scenario, hydrodynamical simulations
show that H\,I gas has a roughly constant covering fraction at radii
$\apl 100\,\rm kpc$ at $z\approx2.5$ \citep[e.g.,][]{Dekel:2009}.  It
is intriguing because the behavior is in a broad agreement with the
observed covering fraction within $\sim 100\,\rm kpc$ in our passive
LRG sample.  However, the cold-mode accretion is expected to be
efficient only for galaxies at high redshifts $z\apg2$
\citep[e.g.,][]{Dekel:2009} or low-mass galaxies at lower redshifts.
At $z<1$, the mechanism might not be effective in high-mass halos, and
cold filaments might be truncated or disrupted before reaching the
center \citep[e.g.,][]{Keres:2009}.
%In addition, the predicted
%metallicity of the cold filaments is likely to be at least an order of
%magnitude lower \citep[$\leq 10^{-3} \,\rm Z_{\odot}$ in][]{Kimm:2011}
%than the metallicity of \ewr $\geq\,0.3\,\rm\AA$ absorbers
%\citep[$\apg 0.03\,\rm Z_{\odot}$ inferred from][]{Murphy:2007}.

In addition to intergalactic gas, satellite accretion is also expected
to proceed along filaments based on the preferential alignment of
satellite galaxies along the major axis of massive central galaxies
(e.g., Yang \etal\ 2006) and the parallel alignment of the major axis
of galaxies and surrounding filaments (e.g., Tempel \etal\ 2013,
2015).  If the observed \MgII absorbers arise in accreted satellites
along a more confined large-scale filament, then we can simultaneously
explain the enhanced $\langle\kappa\rangle_\MgII$ along the major axis
of \OII-emitting LRGs at $d<50$ kpc and the observed reduced velocity
dispersion relative to what is expected for virialized gas in massive
halos.

\section{Summary and Conclusions}

We study the chemically-enriched cool gas content in massive halos
based on a survey of \MgII absorbers associated with 37621 LRGs in the
spectra of background QSOs.  \MgII absorption is detected around both
passive and \OII-emitting LRGs.  The covering fraction of \MgII gas is
higher around \OII-emitting LRGs inside 100 kpc; the covering fraction
in both passive and \OII-emitting LRGs declines rapidly with radius.
Both \MgII-absorbing and non-absorbing LRGs show comparably old
stellar populations. There is a weak azimuthal dependence of
absorption: \OII-emitting LRGs show more absorption along the major
axis.  This trend is only significant within 50 kpc.  The velocity
dispersion of \MgII relative to the LRGs is less than expected for LRG
halo masses in both types of galaxy.

We find that the observed Mg\,II absorbers in the vicinities of LRGs
are best-explained by a combination of cool clouds formed through
thermal instabilities in LRG halos and satellite accretion through
filaments that are preferentially aligned with the major axis of the
LRGs.  While AGN are likely present in nearly all of the LRGs in our
sample, the suppressed velocity dispersion found for the \MgII
absorbing gas makes AGN winds an unlikely contributor.  We expect that
follow-up analysis of available imaging data around the LRGs will
provide the necessary test for the accreting satellite scenario.

\section*{Acknowledgments}

It is a pleasure to thank Jenny Greene, Hung-Jin Huang, Cameron Liang,
Michael Rauch, and Mark Voit for helpful discussions on the
interpretations of our main findings.  We thank an anonymous referee
for constructive comments that helped improve the presentation of the
paper.  HWC acknowledges the Aspen Center for Physics, which is
supported by National Science Foundation grant PHY-1066293, and the
organizers of the workshop on the ``physics of accretion and feedback
in the circumgalactic medium'' for a productive visit in June 2015,
during which components of the work presented were accomplished.  We
are grateful to the SDSS collaboration for producing and maintaining
the SDSS public data archive.  Funding for SDSS-III has been provided
by the Alfred P. Sloan Foundation, the Participating Institutions, the
National Science Foundation, and the U.S. Department of Energy Office
of Science. The SDSS-III web site is http://www.sdss3.org/.

SDSS-III is managed by the Astrophysical Research Consortium for the
Participating Institutions of the SDSS-III Collaboration including the
University of Arizona, the Brazilian Participation Group, Brookhaven
National Laboratory, Carnegie Mellon University, University of
Florida, the French Participation Group, the German Participation
Group, Harvard University, the Instituto de Astrofisica de Canarias,
the Michigan State/Notre Dame/JINA Participation Group, Johns Hopkins
University, Lawrence Berkeley National Laboratory, Max Planck
Institute for Astrophysics, Max Planck Institute for Extraterrestrial
Physics, New Mexico State University, New York University, Ohio State
University, Pennsylvania State University, University of Portsmouth,
Princeton University, the Spanish Participation Group, University of
Tokyo, University of Utah, Vanderbilt University, University of
Virginia, University of Washington, and Yale University.

%\footnotesize{
%\bibliographystyle{mn} 
%\bibliography{./biblio}
%}
%

\label{lastpage}

\end{document}